\documentclass[journal]{IEEEtran}

\usepackage{subfigure}
\usepackage{amsmath,amsfonts,amssymb}
\usepackage[linesnumbered,ruled]{algorithm2e}
\usepackage{array}
\usepackage[caption=false,font=normalsize,labelfont=sf,textfont=sf]{subfig}
\usepackage{textcomp}
\usepackage{stfloats}
\usepackage{url}
\usepackage{verbatim}
\usepackage{graphicx}
\usepackage{cite}
\usepackage{cases}
\usepackage{makecell}

\usepackage{tikz}

\SetKwInput{KwInput}{Inputs}               
\SetKwInput{KwOutput}{Output}              
\SetKwFunction{CalcProbAndSort}{CalcProbAndSort}
\SetKwFunction{GenErrSeq}{GenErrSeq}

\def\C{\mathcal{C}}
\def\L{\mathcal{L}}
\def\S{\mathcal{S}}
\def\preceq{\preccurlyeq}

\newcommand{\overbar}[1]{\mkern 1.5mu\overline{\mkern-1.5mu#1\mkern-1.5mu}\mkern 1.5mu}
\def\phi{\varphi}

\begin{document}

\title{\huge Guessing Random Additive Noise Decoding of Network Coded Data Transmitted over Burst Error Channels}

\author{Ioannis Chatzigeorgiou,~\IEEEmembership{Senior Member,~IEEE,} and Dmitry Savostyanov
\thanks{Part of this paper was presented at the 2022 IEEE International Symposium on Information Theory (ISIT) [DOI: 10.1109/ISIT50566.2022.9834692]. Dmitry Savostyanov acknowledges the support of the Leverhulme Trust Fellowship programme (RF-2021-258). \textit{(Corresponding
author: Ioannis Chatzigeorgiou.)}}
\thanks{Ioannis Chatzigeorgiou is with the School of Computing and Communications, Lancaster University, LA1 4WA, United Kingdom (email: i.chatzigeorgiou@lancaster.ac.uk).}
\thanks{Dmitry Savostyanov is with the School of Mathematics, Statistics and Actuarial Science, University of Essex, CO4 3SQ, United Kingdom (email: d.savostyanov@essex.ac.uk).}
}



\maketitle

\begin{abstract}
We consider a transmitter that encodes data packets using network coding and broadcasts coded packets. A receiver employing network decoding recovers the data packets if a sufficient number of error-free coded packets are gathered. The receiver does not abandon its efforts to recover the data packets if network decoding is unsuccessful; instead, it employs syndrome decoding (SD) in an effort to repair erroneous received coded packets, and then reattempts network decoding. Most decoding techniques, including SD, assume that errors are independently and identically distributed within received coded packets. Motivated by the guessing random additive noise decoding (GRAND) framework, we propose transversal GRAND (T-GRAND): an algorithm that exploits statistical dependence in the occurrence of errors, complements network decoding and recovers all data packets with a higher probability than SD. \mbox{T-GRAND} examines error vectors in order of their likelihood of occurring and altering the transmitted packets. Calculation and sorting of the likelihood values of all error vectors is a simple but computationally expensive process. To reduce the complexity of T-GRAND, we take advantage of the properties of the likelihood function and develop an efficient method, which identifies the most likely error vectors without computing and ordering all likelihood values.
\end{abstract}


\section{Introduction}
\label{sec:intro}
A linear code at the physical layer (PHY) maps a sequence of $K$ symbols onto a sequence of $N>K$ symbols, known as a \textit{codeword}. Whereas bit-level linear codes offer forward error correction at PHY, packet-level linear codes improve the reliability of communication systems at layers above PHY. Once data have been organized into strings of symbols, called \textit{packets}, a packet-level linear code maps a sequence of $K$ data packets onto a sequence of $N>K$ coded packets. When packet-level \textit{random} linear codes (RLCs)~\cite{Ho2006} are used, the $N$ coded packets are random linear combinations of the $K$ data packets. As a result, packet-level random linear coding over the binary field is equivalent to successive column-wise bit-level random linear coding, which means that -- if the $N$ coded packets are stacked to form the rows of a matrix -- the $N$ elements in each column of the matrix~will~compose~a~codeword. 

Packet-level RLCs, henceforth referred to as RLCs, encompass both random linear \textit{network} codes~\cite{Ho2006} and random linear \textit{fountain} codes~\cite{Byers1998}; in the former case, packets are combined at intermediate nodes of a multi-hop network; in the latter case, packets are combined at the input of a single-hop broadcast channel. Whereas automatic repeat request (ARQ) schemes rely extensively on  delay-inducing feedback from receivers to transmitters to support reliable multicast communication, RLCs can increase throughput -- as they require limited or no feedback -- and can be combined with queueing methods to improve network congestion control~\cite{Huang2018}. RLC implementations are integral parts of proposed solutions that use the satellite segment to extend the coverage of terrestrial infrastructures and support passengers on vehicles, high-speed trains and aircraft; examples include the multipath transmission control protocol based on network coding (MPTCP/NC)~\cite{Cloud2013}, fountain code-based MPTCP (FMTCP) \cite{Cui2015} and path-based network coding (PBNC) \cite{Giambene2019}. Furthermore, RLC-based strategies for cooperative coded caching at edge nodes of vehicular networks, e.g., roadside units, can facilitate the flexible and reliable delivery of popular content to connected vehicles \cite{Chen2020}. 

Receivers that employ RLC decoding typically discard coded packets that have been corrupted by errors and attempt to reconstruct data packets from correctly received coded packets only. However, when RLC decoding is unsuccessful, correctly received coded packets can be used in conjunction with erroneous received coded packets. Packetized rateless algebraic consistency (PRAC) \cite{Angelopoulos2014} is an error correction method that exploits the algebraic properties of RLCs to identify codewords that contain errors and iteratively search for valid codewords to replace erroneous codewords. Given the aforementioned row/column correspondence of coded packets and codewords in RLCs, correcting errors in codewords is equivalent to repairing segments in coded packets. Syndrome decoding~\cite{Mohammadi2016} looks at the same problem from a different perspective; if errors are detected in a codeword, syndrome decoding searches for the most likely error vector that corrupted a codeword instead of looking for the most likely valid codeword. Although PRAC and syndrome decoding can improve the chances of an RLC decoder recovering all data packets, they do not account for correlations between errors, i.e., they are more suitable for memoryless channels, like the binary symmetric channel (BSC). To address this operational limitation, we look into the recently proposed guessing random additive noise decoding (GRAND) scheme.

GRAND is a maximum-likelihood decoding scheme that endeavors to identify and nullify the effect of noise on transmitted signals~\cite{Duffy2018,Duffy2019}. In low-noise entropy conditions, guessing the noise is notably faster than performing an exhaustive search through all possible codewords. GRAND has enabled \textit{universal} decoding, as it can decode $\textit{any}$ linear code. Since its inception, GRAND has been extended to non-linear codes~\cite{Cohen2022}, hard-input iterative decoding~\cite{Galligan2021}, soft detection decoding~\cite{Duffy2021} and soft-input soft-output decoding~\cite{Condo2022}. It has also been combined with high-order modulation techniques~\cite{An2022,Chatzigeorgiou2022cl}, modified for fading channels~\cite{Chatzigeorgiou2022cl, Abbas2022, Sarieddeen2022} and implemented in hardware~\cite{Abbas20222vlsi}.

The extension of GRAND to channels with memory was envisioned in~\cite{Duffy2019}. Soon after, GRAND Markov order \mbox{(GRAND-MO)} \cite{An2022} was developed for guessing errors that are correlated over time. GRAND-MO focuses on coding at PHY, according to which binary codewords are transmitted over a burst error channel and, consequently, correlated errors are distributed over the bits of each codeword. By contrast, if binary RLC encoding is used to generate and transmit coded packets over a burst error channel, correlated errors will be distributed over the bits of each coded packet. Remembering that codewords in packet-level RLCs occupy the columns of a matrix formed by the transmitted coded packets as its rows, we deduce that errors \textit{across} adjacent codewords will be correlated but errors \textit{within} each codeword will be uncorrelated. Therefore, GRAND-MO cannot complement RLC decoding at layers above PHY.

Through this literature review, we established that currently available solutions based on GRAND, for channels with or without memory, are specifically designed for PHY. When packet-level RLC is used at layers higher than PHY, syndrome decoding~\cite{Mohammadi2016} is the only available solution that is aligned with the principles of GRAND and complements RLC decoding, provided that the underlying channel is memoryless and, hence, received errors are uncorrelated. The objective of this paper is to propose \textit{transversal GRAND} -- a novel algorithm designed to leverage correlations in errors \textit{across} adjacent codewords, and to work in tandem with RLC decoding at layers higher than PHY. Correlations in errors could be introduced by the communication channel and find their way to layers higher than PHY if stringent latency requirements prevent the use of interleaving \cite{An2022}, or introduced by the channel decoder at the PHY layer -- an event known as \textit{decoder error propagation}~\cite{Zhu2020}.

The basic idea behind transversal GRAND was introduced in our previous work~\cite{Chatzigeorgiou2022isit}, which considered single-hop transmission over a channel with memory. The contributions of this paper are as follows:
\begin{itemize}
\item The algorithmic details of transversal GRAND are described extensively and examples are provided to illustrate how knowledge of the statistical description of the burst error process can improve the chances of erroneous received coded packets being repaired.
\item The computational complexity of a core component of transversal GRAND is studied thoroughly and its efficiency is discussed.
\item A novel approach of adjustable computational complexity is developed and contrasted with the original implementation of the core component of transversal GRAND.
\item Simulations of coded packet transmissions over a burst error channel verify and quantify the advantages of transversal GRAND over syndrome decoding. Numerical results also demonstrate that the computational complexity of transversal GRAND can be reduced without negatively impacting its performance.
\end{itemize}

The remainder of this paper has been organized as follows: Section~\ref{sec:RLC} describes RLC decoding, which discards packets received in error. Section~\ref{sec:SD} explains how syndrome decoding can repair packets in error and assist RLC decoding. Transversal GRAND is introduced in Section~\ref{sec:TGRAND} as a means to exploit the statistical structure of errors and improve the chances of repairing packets. The implementation of a core procedure of transversal GRAND, which is responsible for sorting error vectors in order of likelihood, is discussed in Section~\ref{sec:Sorting} and revised in Section~\ref{sec:Tracing} to reduce its complexity. Section~\ref{sec:Comp_char} dwells deeper into computational aspects of the proposed sorting procedures. Performance comparisons between syndrome decoding and transversal GRAND are presented in Section~\ref{sec:Results} and key findings are summarized in Section~\ref{sec:Conclusions}.


\section{Packet-Level Random Linear Coding\\and Decoding}
\label{sec:RLC}

We consider a source node, which is about to broadcast messages to one or more destination nodes. Before transmission, each message is segmented into $K$ source packets. Each source packet has been modeled as a sequence of $B$ bits. The $K$ source packets of $B$ bits can be expressed as a matrix $\mathbf{U}\in\mathbb{F}^{K\times B}_{2}$, where $\mathbb{F}^{K\times B}_{2}$ denotes the set of all $K\times B$ matrices over $\mathbb{F}_{2}=\{0,1\}$. Random linear network coding~\cite{Ho2006} is used to encode the $K$ source packets into $N\geq K$ coded packets. The $N$ coded packets can also be expressed in matrix form as $\mathbf{X}\in\mathbb{F}^{N\times B}_{2}$. The relationship between matrices $\mathbf{X}$ and $\mathbf{U}$ is:
\begin{equation}
\label{eq:RLC}
\mathbf{X} = \mathbf{G}\,\mathbf{U},
\end{equation}
where $\mathbf{G}$ is known as the $N\times K$ \textit{generator matrix} of the RLC. We employ \textit{systematic} RLC encoding, according to which the first $K$ of the $N$ transmitted packets are identical to the $K$ source packets, while the remaining $N-K$ coded packets are random linear combinations of the source packets. The generator matrix $\mathbf{G}$ can thus be expressed as:
\begin{equation}
\label{eq:generator_matrix_structure}
\mathbf{G}=
\left[\!\begin{array}{c} \mathbf{I}_K\\ \mathbf{P}\end{array}\!\right],
\end{equation}
where $\mathbf{I}_K$ is the $K\times K$ identity matrix. Each element of the $(N-K)\times K$ matrix $\mathbf{P}$ is chosen uniformly and at random from $\mathbb{F}_2$. The value of the seed that initializes the pseudo-random number generator and contributes to the construction of $\mathbf{P}$ at the source node can be conveyed in the headers of the coded packets, as explained in \cite{Jin2008}. For this reason, we assume that destination nodes have knowledge of $\mathbf{P}$.

Let $\mathbf{Y}$ be an erroneous copy of $\mathbf{X}$ that has been received by a destination node. The destination node classifies the received coded packets into error-free and erroneous coded packets, e.g., using cyclic redundancy checks (CRCs). We denote by $\mathcal{R}$ the set of the row indices of $\mathbf{Y}$ that correspond to correctly received coded packets, and by $N_\mathrm{R}$ the number of coded packets that contain no errors, i.e., $\vert \mathcal{R}\vert=N_\mathrm{R}\leq N$. The remaining $N-N_\mathrm{R}$ received coded packets, which have been corrupted by errors, have indices in the set $\overbar{\mathcal{R}}=\{1,\ldots,N\}\backslash \mathcal{R}$. The destination node constructs matrices $\mathbf{Y}_\mathcal{R}$ and $\mathbf{G}_\mathcal{R}$ from those $N_\mathrm{R}$ rows of $\mathbf{Y}$ and $\mathbf{G}$, respectively, with indices in $\mathcal{R}$. On the other hand, the indices in $\overbar{\mathcal{R}}$ identify the $N-N_\mathrm{R}$ rows of $\mathbf{Y}$ that will be used in the construction of matrix $\mathbf{Y}_{\overbar{\mathcal{R}}}$. Matrices $\mathbf{Y}$, $\mathbf{Y}_\mathcal{R}$ and $\mathbf{Y}_{\overbar{\mathcal{R}}}$ can be expressed as follows:
\begin{subnumcases}
{\mathbf{Y} = \mathbf{X}\oplus\mathbf{E}\Leftrightarrow \label{eq:received_matrix}}
      \mathbf{Y}_\mathcal{R}=\mathbf{X}_\mathcal{R}=\mathbf{G}_\mathcal{R}\mathbf{U},\label{eq:error_free}\\
      \mathbf{Y}_{\overbar{\mathcal{R}}}=\mathbf{X}_{\overbar{\mathcal{R}}}\oplus\mathbf{E}_{\overbar{\mathcal{R}}},\label{eq:erroneous}
\end{subnumcases}
where $\oplus$ represents modulo-$2$ addition. The \textit{error matrix} $\mathbf{E}$ contains ones in positions where errors have occurred and zeros in the remaining positions. The indices in sets $\mathcal{R}$ and $\overbar{\mathcal{R}}$ can be used to decompose $\mathbf{E}$ into $\mathbf{E}_{\mathcal{R}}$ and $\mathbf{E}_{\overbar{\mathcal{R}}}$, where $\mathbf{E}_{\mathcal{R}}$ is equal to the $N_\mathrm{R}\times B$ zero matrix.

Decoding of an RLC utilizes the $N_\mathrm{R}$ correctly received coded packets, represented by $\mathbf{Y}_\mathcal{R}$, and discards the $N-N_\mathrm{R}$ erroneous received coded packets contained in $\mathbf{Y}_{\overbar{\mathcal{R}}}$. The source message, represented by $\mathbf{U}$, can be obtained from $\mathbf{X}_\mathcal{R} = \mathbf{G}_\mathcal{R}\,\mathbf{U}$ in \eqref{eq:error_free} if the rank of $\mathbf{G}_\mathcal{R}$ is $K$. In that case, the relationship $\mathbf{X}_\mathcal{R} = \mathbf{G}_\mathcal{R}\,\mathbf{U}$ can be seen as a system of $N_\mathrm{R}\geq K$ linear equations, which can be reduced to a system of $K$ linearly independent equations with $K$ unknowns, i.e., source packets. This $K\times K$ system will return a unique solution for $\mathbf{U}$. Otherwise, if $\mathrm{rank}(\mathbf{G}_\mathcal{R})<K$, the linear system is underdetermined, i.e., it has fewer linearly independent equations than unknowns, therefore a unique solution cannot be obtained.

If the RLC decoder does not discard the erroneous coded packets but attempts to repair them, the probability of recovering the source message could potentially be increased. The following section describes syndrome decoding~\cite{Mohammadi2016}, which aims to repair erroneous coded packets without taking into consideration the statistical properties of the channel.


\section{Guessing Uncorrelated Errors:\\Syndrome Decoding}
\label{sec:SD}

We established in \eqref{eq:erroneous} that the $N-N_\mathrm{R}$ erroneous coded packets, which compose $\mathbf{Y}_{\overbar{\mathcal{R}}}$, can be expressed as the \mbox{modulo-$2$} sum of the corresponding transmitted coded packets, which form the rows of $\mathbf{X}_{\overbar{\mathcal{R}}}$, and matrix $\mathbf{E}_{\overbar{\mathcal{R}}}$. If an estimate of $\mathbf{E}_{\overbar{\mathcal{R}}}$, denoted by $\hat{\mathbf{E}}_{\overbar{\mathcal{R}}}$, is computed, an estimate of the transmitted coded packets can be derived as follows:
\begin{equation}
\label{eq:Estimated_X}
\hat{\mathbf{X}}_{\overbar{\mathcal{R}}}=\mathbf{Y}_{\overbar{\mathcal{R}}}\oplus\hat{\mathbf{E}}_{\overbar{\mathcal{R}}}.
\end{equation}
CRC verification will determine which of the estimated rows of $\hat{\mathbf{X}}_{\overbar{\mathcal{R}}}$ correspond to successfully repaired coded packets. Let $\nu$ denote the number of rows in $\hat{\mathbf{X}}_{\overbar{\mathcal{R}}}$ that passed CRC verification. The indices of the $\nu$ successfully repaired coded packets will be removed from set $\overbar{\mathcal{R}}$ and will be added to set $\mathcal{R}$, while the corresponding rows of $\hat{\mathbf{X}}_{\overbar{\mathcal{R}}}$ will be moved to $\mathbf{Y}_\mathcal{R}$ in \eqref{eq:error_free}. The cardinalities of sets $\overbar{\mathcal{R}}$ and $\mathcal{R}$ will change to $N-N_\mathrm{R}-\nu$ and $N_\mathrm{R}+\nu$, respectively, while the dimensions of $\hat{\mathbf{X}}_{\overbar{\mathcal{R}}}$ and $\mathbf{Y}_\mathcal{R}$ will change to $(N-N_\mathrm{R}-\nu)\times B$ and $(N_\mathrm{R}+\nu)\times B$, respectively. The indices of the $\nu$ repaired packets will also be used to identify the rows of the generator matrix $\mathbf{G}$ that should be appended to $\mathbf{G}_\mathcal{R}$ in \eqref{eq:error_free}. If the $\nu$ repaired packets increase the rank of the enlarged $(N_\mathrm{R}+\nu)\times K$ matrix $\mathbf{G}_\mathcal{R}$ to $K$, the process of estimating, or `guessing', matrix $\mathbf{E}_{\overbar{\mathcal{R}}}$ will have been successful in assisting the RLC decoder to recover the source message.

Syndrome decoding is a method proposed in \cite{Mohammadi2016} for the calculation of $\hat{\mathbf{E}}_{\overbar{\mathcal{R}}}$. This method takes into account that the \mbox{$(N-K)\times K$} matrix $\mathbf{P}$ matrix $\mathbf{P}$ is known to all destination nodes, and derives the $N\times (N-K)$ \textit{parity-check matrix} $\mathbf{H}$ as follows:
\begin{equation}
\label{eq:sys_parity_check_matrix}
\mathbf{H}=\left[\!\begin{array}{c} -\mathbf{P}\;\vert\;\mathbf{I}_{N-K}\end{array}\!\right]^\top,
\end{equation}
so that:
\begin{equation}
\label{eq:zero_product}
\mathbf{H}^\top\,\mathbf{G}=\mathbf{0}.
\end{equation}
When operations are performed in $\mathbb{F}_2$, negation has no effect on a matrix, that is, $-\mathbf{P}=\mathbf{P}$. Multiplication of $\mathbf{H}^\top$ by the received matrix $\mathbf{Y}$ produces the $(N-K)\times B$ \textit{syndrome matrix}~$\mathbf{S}$, i.e., $\mathbf{S} = \mathbf{H}^\top \mathbf{Y}$. Using \eqref{eq:RLC}, \eqref{eq:received_matrix} and \eqref{eq:zero_product}, we find that the relationship between the syndrome matrix $\mathbf{S}$ and the error matrix $\mathbf{E}$ is:
\begin{equation}
\label{eq:syndrome_decoding_full}
\mathbf{S} = \mathbf{H}^\top \mathbf{Y}
 = \mathbf{H}^\top (\mathbf{X}\oplus\mathbf{E})
 = \mathbf{H}^\top (\mathbf{G}\mathbf{U}\oplus\mathbf{E})
 = \mathbf{H}^\top \mathbf{E}.
\end{equation}
As our focus is on the $N-N_\mathrm{R}$ erroneous coded packets only, we use the set $\overbar{\mathcal{R}}$ to isolate the $N-N_\mathrm{R}$ of the $N$ rows of $\mathbf{H}$ and $\mathbf{E}$, and reduce the two matrices to $\mathbf{H}_{\overbar{\mathcal{R}}}$ and $\mathbf{E}_{\overbar{\mathcal{R}}}$, respectively. Consequently, expression \eqref{eq:syndrome_decoding_full} changes to:
\begin{equation}
\label{eq:syndrome_decoding_partial}
\mathbf{S} = \left(\mathbf{H}_{\overbar{\mathcal{R}}}\right)^{\!\top} \mathbf{E}_{\overbar{\mathcal{R}}}.
\end{equation}
If column $b$ of $\mathbf{S}$ and $\mathbf{E}_{\overbar{\mathcal{R}}}$ is denoted by $[\mathbf{S}]_{*,b}$ and $[\mathbf{E}_{\overbar{\mathcal{R}}}]_{*,b}$, respectively, expression \eqref{eq:syndrome_decoding_partial} can be re-written as $B$ independent systems of $N-K$ linear equations with $N-N_\mathrm{R}$ unknowns per equation:
\begin{equation}
\label{eq:syndrome_decoding_partial_per_col}
\left[\mathbf{S}\right]_{*,b} = \left(\mathbf{H}_{\overbar{\mathcal{R}}}\right)^{\!\top} \left[\mathbf{E}_{\overbar{\mathcal{R}}}\right]_{*,b}\quad\text{for}\quad b=1,\ldots,B.
\end{equation}

Mohammadi~\textit{et al.}~\cite{Mohammadi2016} observed that erroneous received coded packets usually contain large error-free segments, thus $\mathbf{E}_{\overbar{\mathcal{R}}}$ is a \textit{sparse} matrix, that is, most of the elements in $\mathbf{E}_{\overbar{\mathcal{R}}}$ are zero-valued. Based on this observation, the solution to \eqref{eq:syndrome_decoding_partial_per_col} can be formulated as:
\begin{IEEEeqnarray}{ll}
\label{eq:norm_minimization}
\bigl[\hat{\mathbf{E}}_{\overbar{\mathcal{R}}}\bigr]_{*,b}=\;  &\arg\min_{\mathbf{w}^{\top}}\;\lVert \mathbf{w} \rVert_0 \IEEEyesnumber\IEEEyessubnumber*\label{eq:obj_func}\\
&\text{subject to}\; \left(\mathbf{H}_{\overbar{\mathcal{R}}}\right)^{\!\top}\mathbf{w}^{\top}=\left[\mathbf{S}\right]_{*,b}
\label{eq:constraint}
\end{IEEEeqnarray}
where $\mathbf{w}\in\mathbb{F}^{N-N_\mathrm{R}}_2$ is a row vector that satisfies constraint \eqref{eq:constraint} and has the minimum possible number of non-zero elements. The norm $\lVert \mathbf{w} \rVert_0$, which counts the non-zero elements in $\mathbf{w}$, is defined as $\lVert \mathbf{w} \rVert_0=\lvert w_1\rvert^0+\ldots+\lvert w_{N-N_\mathrm{R}}\rvert^0$ assuming that $0^0=0$ \cite{Donoho2001}. Syndrome decoding considers \eqref{eq:obj_func} and initiates an exhaustive search for a solution; the sparsity of the row vector $\mathbf{w}$ is gradually reduced, i.e., the number of ones in $\mathbf{w}$ increases, and the search concludes when the sparsest vector $\mathbf{w}$ that satisfies \eqref{eq:constraint} has been identified.

Although syndrome decoding has the potential to assist RLC decoding and increase the probability of recovering the source packets~\cite{Mohammadi2016}, it does not consider the statistical properties of the channel. It selects columns for $\hat{\mathbf{E}}_{\overbar{\mathcal{R}}}$ that are as sparse as possible or, equivalently, have the lowest possible Hamming weight, but it does not account for the possibility that the columns of $\hat{\mathbf{E}}_{\overbar{\mathcal{R}}}$ are correlated. Thus, syndrome decoding is suitable for memoryless channels only, such as the BSC. The following section describes a simplified version of the Gilbert-Elliott channel model, which is commonly used for the characterization of burst errors in channels with memory, and introduces transversal GRAND, which exploits the statistical properties of burst error channels.


\section{Guessing Burst Errors: Transversal GRAND}
\label{sec:TGRAND}

Coded packets at the input of the RLC decoder often contain bit errors, the distribution of which is governed by the channel type, multi-user interference, the modulation method and the forward error correction scheme at the physical layer. The wireless transmission medium and the physical layer at the source and destination nodes form a composite channel, which introduces time-dependent correlated error bursts. Gilbert~\cite{Gilbert1960} modeled burst error channels, and Elliott~\cite{Elliott1963} proposed the more general Gilbert-Elliott channel model. The analysis of RLC-based schemes over channels with memory frequently relies on the Gilbert-Elliott channel to model transmission over wireless channels, e.g., \cite{Cloud2015,Jun2017,Huang2021}.

\begin{figure}[t]
\centering
\includegraphics[width=0.5\columnwidth]{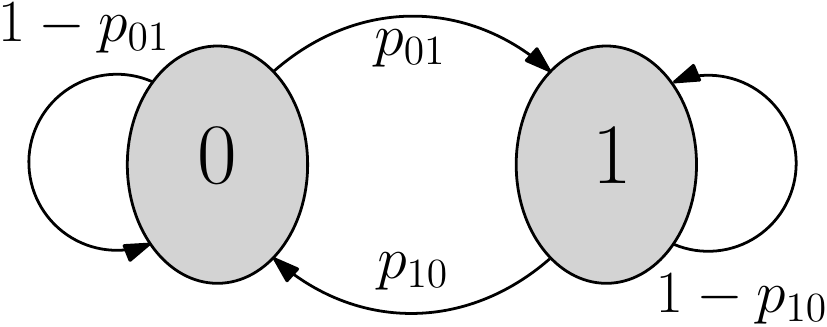}
\caption{The simplified Gilbert-Elliott channel model, where $0$ and $1$ represent the `good' state and the `bad' state, respectively.}
\label{fig:channel_model}
\end{figure}

\subsection{The Simplified Gilbert-Elliott Channel Model}
\label{sec:Gilbert}

The Gilbert-Elliott channel model is a two-state Markov chain. The channel state can be in a `bad' state or a `good' state, and can either transition from one state to the other or remain in the same state. In the bad state, a bit error is generated with a certain high probability, while in the good state a bit error is generated with a low probability. A simplification of the model, which is often considered in the literature, e.g., \cite{Aboutorab2014,Cideciyan2019, An2022}, assumes that the bad state always induces a bit error, while the good state never flips the value of a bit. Therefore, the simplified channel model can be reduced to the Markov chain shown in Fig.~\ref{fig:channel_model}, where the bad state is denoted by $1$ and the good state is denoted by $0$. A transition from state~$0$ to state~$1$ occurs with probability $p_{01}$ and causes a transmitted bit to be received in error. A transition from state~$1$ to state~$0$, which occurs with probability $p_{10}$, implies that a transmitted bit has been received correctly. The channel may remain in state~$0$ or state $1$ with probability $1-p_{01}$ or $1-p_{10}$, respectively.

The steady-state probability of being in state $1$ represents the bit error probability, which is given by $\varepsilon=p_{01}/(p_{01}+p_{10})$ \cite{Rancurel2001,Cloud2015}. The expected number of consecutive errors provides the average length of an error burst, and is equal to $\Lambda=1/p_{10}$ \cite{Cloud2015}. The memory of a burst error channel is defined as $\mu=1-p_{10}-p_{01}$~\cite{Mushkin1989} and provides a measure of the persistence of the channel in remaining in a state. If $-1\leq\mu<0$, the channel has oscillatory memory. Otherwise, if $0<\mu<1$, the channel has persistent memory. For $\mu=0$, the channel is memoryless.

Estimation of the bit error probability $\varepsilon$ and the average burst length $\Lambda$ facilitates computation of the transition probabilities $p_{01}$ and $p_{10}$, as the latter can be expressed in terms of the former, that is, $p_{01}=\varepsilon/\left[\Lambda\left(1-\varepsilon\right)\right]$ and $p_{10}=1/\Lambda$. The values of $p_{01}$ and $p_{10}$ can then be used as inputs to our proposed algorithm and guide the derivation of matrix $\hat{\mathbf{E}}_{\overbar{\mathcal{R}}}$.

\subsection{The Transversal GRAND Algorithm}
\label{subsec:focus_on_TGRAND}

Let us refer to a prospective column of $\hat{\mathbf{E}}_{\overbar{\mathcal{R}}}$ as an \textit{error vector} of length $L=N-N_\mathrm{R}$. In syndrome decoding, error vectors that have the same number of non-zero elements, i.e., the same Hamming weight, are treated as being equally likely and can thus be queried in any order by \eqref{eq:norm_minimization}. Furthermore, all error vectors of a given Hamming weight $w$ are queried before error vectors of weight $w+1$. The proposed method, which we call transversal GRAND, queries error vectors in descending order of likelihood, as dictated by the transition probabilities of the channel model and not by the Hamming weights of the error vectors.

\begin{algorithm}[t]
\label{alg:TGRAND1}
\caption{Implementation of the transversal GRAND algorithm according to which generation of the sequence of error vectors relies on the calculation and sorting of the probabilities of occurrence of the vectors in descending order of magnitude.}
\DontPrintSemicolon
	\KwInput{$\mathbf{H}_{\overbar{\mathcal{R}}}$, $\mathbf{S}$, $p_{01}$, $p_{10}$}
  	\KwOutput{$\hat{\mathbf{E}}_{\overbar{\mathcal{R}}}$}
  	\textit{origin\textunderscore vector} $\longleftarrow\mathbf{0}$\label{line:TGRAND1_initialise}
  	
	\For{$b\leftarrow 1$ \KwTo $B$}
  	{
  		\label{line:TGRAND1_forall_columns}
            Determine $L_0$ and $L_1$ from \textit{origin\textunderscore vector} \label{line:TGRAND1_L0_L1}
            
            \tcp{\textit{origin\textunderscore vector} changes with $b$}

    		$\bigl\{(\ell^{(\ell)}_0, \ell^{(\ell)}_1)\bigr\}_{\ell=1}^{(L_0+1)(L_1+1)}\longleftarrow$ \phantom{--------}~\CalcProbAndSort{$p_{01}$, $p_{10}$, $L_0$, $L_1$} \label{line:TGRAND1_sort}\;    		
		\For{$\ell\leftarrow 1$ \KwTo $(L_0+1)(L_1+1)$}
    		{
			\label{line:TGRAND1_forall_prob}
        			\textit{err\textunderscore seq}$\;\longleftarrow$ \GenErrSeq{$\ell^{(\ell)}_0$, $\ell^{(\ell)}_1$, $L_0$, $L_1$} \label{line:TGRAND1_error_seq}\;
        			
			\ForAll{error vectors in err\textunderscore seq}
        			{
				\label{line:TGRAND1_query}
            			$\mathbf{w}^{\top}\longleftarrow$ error vector from \textit{err\textunderscore seq}\;
            			\If{$\left(\mathbf{H}_{\overbar{\mathcal{R}}}\right)^{\!\top}\mathbf{w}^{\top}=\left[\mathbf{S}\right]_{*,b}$}
            			{
                				$\bigl[\hat{\mathbf{E}}_{\overbar{\mathcal{R}}}\bigr]_{*,b}\longleftarrow \mathbf{w}^{\top}$\;
                				\textit{origin\textunderscore vector} $\longleftarrow \bigl[\hat{\mathbf{E}}_{\overbar{\mathcal{R}}}\bigr]_{*,b}$\;
                				\textbf{goto} 16
            			}
        			}
    		}\label{line:TGRAND1_end_forall_prob}
  	}
  	\Return $\hat{\mathbf{E}}_{\overbar{\mathcal{R}}}$ \label{line:TGRAND1_return}\;
\end{algorithm}

As explained in Algorithm~\ref{alg:TGRAND1}, transversal GRAND uses the reduced parity-check matrix $\mathbf{H}_{\overbar{\mathcal{R}}}$, the syndrome matrix $\mathbf{S}$ and the transition probabilities $p_{01}$ and $p_{10}$ of the channel model to compute $\hat{\mathbf{E}}_{\overbar{\mathcal{R}}}$. Calculation of column $b$ of $\hat{\mathbf{E}}_{\overbar{\mathcal{R}}}$ starts with the definition of an `origin' vector of length $L$, which is set equal to column $b-1$ of $\hat{\mathbf{E}}_{\overbar{\mathcal{R}}}$ when column $b$ is being derived, for \mbox{$b=1,\ldots,B$}. An all-zero origin vector is considered in the calculation of the first column of $\hat{\mathbf{E}}_{\overbar{\mathcal{R}}}$ (see line~\ref{line:TGRAND1_initialise} of Algorithm~\ref{alg:TGRAND1}). Essentially, the origin vector contains the current states of $L$ independent Markov chains, i.e., one Markov chain for each row of $\hat{\mathbf{E}}_{\overbar{\mathcal{R}}}$. Initializing the origin vector to the all-zero vector for $b=1$ is equivalent to setting the initial state of each of the $L$ chains to $0$. In general, for any \mbox{$b\in\{1,B\}$}, the origin vector will contain $L_0$ zeros and $L_1$ ones, that is, $L_0$ of the chains will be in state $0$ and the remaining $L_1$ chains will be in state $1$, where $L_0+L_1=L$ (see line \ref{line:TGRAND1_L0_L1} of Algorithm~\ref{alg:TGRAND1}).

According to the channel model, the values of $\ell_i$ of the $L_i$ elements of the origin vector will change from $i$ to $j$ with probability $p_{ij}$, for $i,j\in\{0,1\}$, $i\neq j$ and $\ell_i=0,\dots,L_i$. The values of the remaining $L_i-\ell_i$ elements will stay the same with probability $1-p_{ij}$. The overall probability of occurrence of the aforementioned transitions is:
\begin{equation}
\label{eq:prob_occurence}
f(\ell_0,\ell_1)=p_{01}^{\ell_0}\left(1-p_{01}\right)^{L_0-\ell_0}p_{10}^{\ell_1}\left(1-p_{10}\right)^{L_1-\ell_1}.
\end{equation}
The probability of occurrence $f(\ell_0,\ell_1)$ is calculated for $\ell_0=0,\ldots,L_0$ and $\ell_1=0,\ldots,L_1$, and the $(L_0+1)(L_1+1)$ probability values compose the entries of the $(L_0+1) \times (L_1+1)$ matrix $\mathbf{F}=\begin{bmatrix} f(\ell_0,\ell_1) \end{bmatrix}$. The entries of matrix $\mathbf{F}$ are then sorted in descending order. The coordinates of the sorted entries constitute the set $\{(\ell_0^{(\ell)},\ell_1^{(\ell)})\}_{\ell=1}^{(L_0+1)(L_1+1)}$, as shown in line~\ref{line:TGRAND1_sort} of Algorithm~\ref{alg:TGRAND1}, where index $\ell$ indicates the position of the probability value in the sequence of sorted entries:
\begin{equation}
\label{eq:sequece_of_prob_occurence}
\begin{split}
f(\ell^{(1)}_0,\ell^{(1)}_1)&\geq f(\ell^{(2)}_0,\ell^{(2)}_1)\geq \ldots\\
\ldots&\geq f(\ell^{((L_0+1)(L_1+1))}_0,\ell^{((L_0+1)(L_1+1))}_1).
\end{split}
\end{equation}

For each coordinate pair $(\ell_0^{(\ell)},\ell_1^{(\ell)})$, the algorithm generates a sequence of $\binom{L_0}{\ell^{(\ell)}_0}\binom{L_1}{\ell^{(\ell)}_1}$ equally likely error vectors (see line \ref{line:TGRAND1_error_seq} of Algorithm~\ref{alg:TGRAND1}). The algorithm sequentially considers each error vector in the sequence, assigns it to $\mathbf{w}^\top$ and checks whether $\mathbf{w}^\top$ is a solution to the linear equation $\left(\mathbf{H}_{\overbar{\mathcal{R}}}\right)^{\!\top}\mathbf{w}^{\top}=\left[\mathbf{S}\right]_{*,b}$. If $\mathbf{w}^\top$ is indeed a solution, then it is selected to be column $b$ of $\hat{\mathbf{E}}_{\overbar{\mathcal{R}}}$. Otherwise, the algorithm continues to generate and query error vectors in descending order of likelihood, until a solution is found (see lines \ref{line:TGRAND1_forall_prob}--\ref{line:TGRAND1_end_forall_prob} of Algorithm~\ref{alg:TGRAND1}). As expected, the $(L_0+1)(L_1+1)$ probability values are mapped onto a total of
\begin{equation}
\begin{split}
\sum_{\ell_0=0}^{L_0}\sum_{\ell_1=0}^{L_1}\binom{L_0}{\ell_0}\binom{L_1}{\ell_1}&=\sum_{\ell_0=0}^{L_0}\binom{L_0}{\ell_0}\,\sum_{\ell_1=0}^{L_1}\binom{L_1}{\ell_1}\\
&=2^{L_0}\,2^{L_1}=2^{L_0+L_1}=2^L\nonumber
\end{split}
\end{equation}
error vectors. Transversal GRAND terminates when solutions for all $B$ linear equations have been estimated and $\hat{\mathbf{E}}_{\overbar{\mathcal{R}}}$ has been obtained (see line~\ref{line:TGRAND1_return} of Algorithm~\ref{alg:TGRAND1}).

\begin{figure}[t]
\centering
\includegraphics[width=0.65\columnwidth]{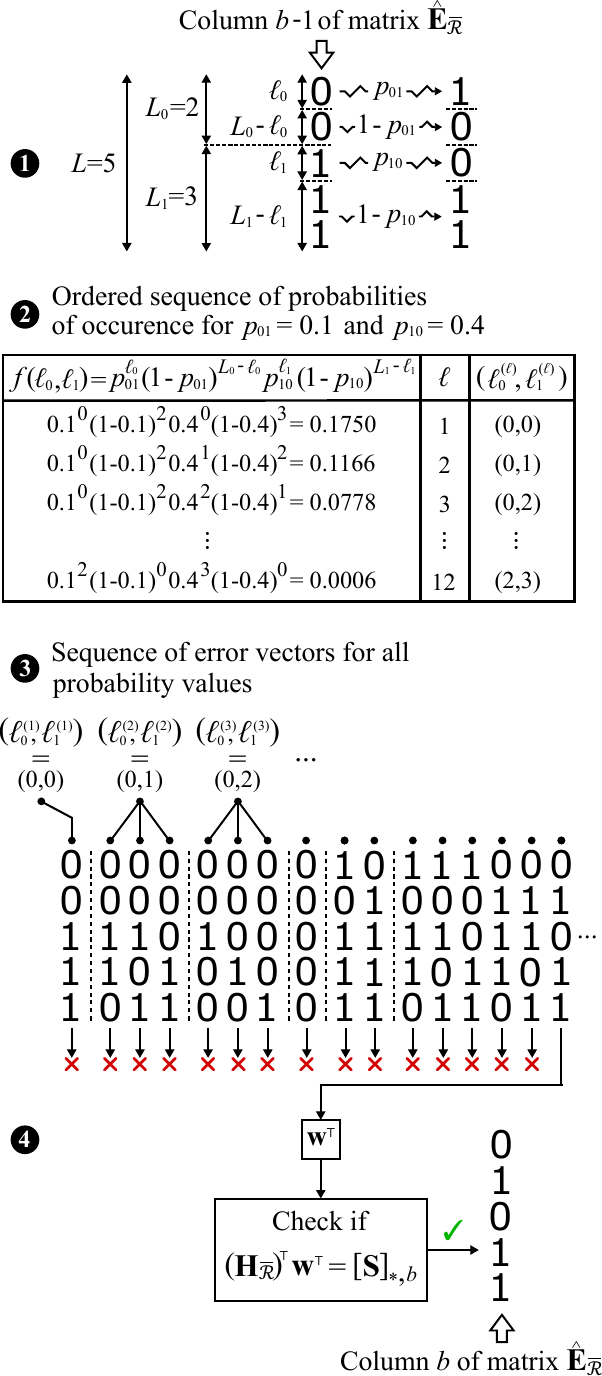}
\caption{Example of the derivation of column $b$ of $\hat{\mathbf{E}}_{\overbar{\mathcal{R}}}$ using Algorithm~\ref{alg:TGRAND1} when the previous column of $\hat{\mathbf{E}}_{\overbar{\mathcal{R}}}$ contains $L_0=2$ zeros and $L_1=3$ ones. The transition probabilities are set to $p_{01}=0.1$ and $p_{10}=0.4$.}
\label{fig:transversal_GRAND1}
\end{figure}

Fig.~\ref{fig:transversal_GRAND1} depicts an example of how transversal GRAND derives column $b$ of $\hat{\mathbf{E}}_{\overbar{\mathcal{R}}}$ from column $b-1$, when $\hat{\mathbf{E}}_{\overbar{\mathcal{R}}}$ consists of $L=5$ rows and the transition probabilities of the channel model are $p_{01}=0.1$ and $p_{10}=0.4$. Notice in Fig.~\ref{fig:transversal_GRAND1} that column $b-1$ of $\hat{\mathbf{E}}_{\overbar{\mathcal{R}}}$, which is represented by the origin vector in Algorithm~\ref{alg:TGRAND1}, contains $L_0=2$ zeros and $L_1=3$ ones. Expression~\eqref{eq:prob_occurence} is used to calculate the probability of an error vector occurring after column $b-1$ of $\hat{\mathbf{E}}_{\overbar{\mathcal{R}}}$ has been observed. Of the $L_0=2$ zeros in column $b-1$ of $\hat{\mathbf{E}}$, $\ell_0$ zeros could change to $1$ with probability $p_{01}$, while the remaining $L_0-\ell_0$ zeros would not change value with probability $1-p_{01}$, for $\ell_0=0,1,2$. On the other hand, $\ell_1$ of the $L_1=3$ ones could change to $0$ with probability $p_{10}$ and the remaining $L_1-\ell_1$ ones would stay unchanged with probability $1-p_{10}$, for $\ell_1=0,1,2,3$. A total of $(L_0+1)(L_1+1)=12$ probability values are computed, stored in the $3\times 4$ matrix~$\mathbf{F}$, arranged in descending order and assigned an ascending index $\ell$. The values of $\ell_0$ and $\ell_1$ that result in the $\ell$-th highest probability value are denoted as $(\ell_0^{(\ell)},\ell_1^{(\ell)})$, as shown in Fig.~\ref{fig:transversal_GRAND1}. For instance, the ordered sequence of probabilities of occurrence in Fig.~\ref{fig:transversal_GRAND1} conveys that the probability of $\ell_0=0$ zeros changing to $1$ and $\ell_1=2$ ones changing to $0$ is $0.0778$, which is the third highest probability value; hence, $(\ell_0^{(3)},\ell_1^{(3)})=(0,2)$. This means that, when $\ell=3$, the algorithm generates and queries $\binom{2}{0}\binom{3}{2}=3$ error vectors -- each of which occurs with probability $0.0778$ -- for column $b$ of $\hat{\mathbf{E}}_{\overbar{\mathcal{R}}}$. Error vectors are created in order of likelihood, as dictated by the ordered sequence of probabilities of occurrence, and each vector is assigned to $\mathbf{w}^{\top}$. The first error vector that satisfies $\left(\mathbf{H}_{\overbar{\mathcal{R}}}\right)^{\!\top}\mathbf{w}^{\top}=\left[\mathbf{S}\right]_{*,b}$ is selected to be column $b$ of $\hat{\mathbf{E}}_{\overbar{\mathcal{R}}}$.

In contrast to syndrome decoding, the error vectors that are generated by transversal GRAND are not necessarily of increasing Hamming weight. Furthermore, transversal GRAND does not assign by default the same probability of occurrence to error vectors that have the same Hamming weight. For example, notice that the leading weight-$3$ error vector in the sequence of error vectors shown in Fig.~\ref{fig:transversal_GRAND1} is not followed by error vectors of higher weights and is not clustered together with other weight-$3$ error vectors.

To conclude this section, we should note that transversal GRAND -- as described by Algorithm~\ref{alg:TGRAND1} -- solves an optimization problem analogous to \eqref{eq:norm_minimization} for syndrome decoding, which can be formulated as follows:
\begin{IEEEeqnarray}{ll}
\label{eq:occur_prob_maximization}
\bigl[\hat{\mathbf{E}}_{\overbar{\mathcal{R}}}\bigr]_{*,b}=\;  &\arg\max_{\mathbf{w}^{\top}}\;f(\ell_0,\ell_1)  \IEEEyesnumber\IEEEyessubnumber*\label{eq:new_obj_func}\\
&\text{where}\; \ell_0 = \lVert \overline{\bigl[\hat{\mathbf{E}}_{\overbar{\mathcal{R}}}\bigr]}_{*,b-1} \odot \mathbf{w}^{\top} \rVert_0\label{eq:compute_ell0}\\
&\phantom{\text{where}}\; \ell_1 = \lVert \bigl[\hat{\mathbf{E}}_{\overbar{\mathcal{R}}}\bigr]_{*,b-1} \odot \overline{\mathbf{w}}^{\top} \rVert_0\label{eq:compute_ell1}\\
&\text{subject to}\; \left(\mathbf{H}_{\overbar{\mathcal{R}}}\right)^{\!\top}\mathbf{w}^{\top}=\left[\mathbf{S}\right]_{*,b}\label{eq:same_constraint}
\end{IEEEeqnarray}
where $\odot$ represents element-wise multiplication, known as the Hadamard product, and $\overbar{\mathbf{v}}$ denotes the ones' complement of a binary vector $\mathbf{v}$. For $b=1$, in accordance with Algorithm~\ref{alg:TGRAND1}, we set $\ell_0=\lVert \mathbf{w} \rVert_0$ and $\ell_1=0$. Transversal GRAND selects a vector $\mathbf{w}^{\top}$ for column $b$ of 
$\hat{\mathbf{E}}_{\overbar{\mathcal{R}}}$ if it maximizes the probability of occurrence $f(\ell_0,\ell_1)$, as shown in \eqref{eq:new_obj_func}, and satisfies \eqref{eq:same_constraint}; the values of $\ell_0$ and $\ell_1$ can be obtained from \eqref{eq:compute_ell0} and \eqref{eq:compute_ell1}, respectively, using the already estimated column $b-1$ of $\hat{\mathbf{E}}_{\overbar{\mathcal{R}}}$ and the vector under consideration.


\section{Implementation of the Sorting Procedure}
\label{sec:Sorting}

A core element of transversal GRAND is the calculation of the probability of occurrence of every possible error vector that could form column $b$ of $\hat{\mathbf{E}}_{\overbar{\mathcal{R}}}$, given column $b-1$ and the transition probabilities $p_{01}$ and $p_{10}$. As explained in Section~\ref{subsec:focus_on_TGRAND}, the probability of occurrence~$f(\ell_0, \ell_1)$ is calculated for $\ell_0=0,\ldots,L_0$ and $\ell_1=0,\ldots,L_1$, where the values of $L_0$ and $L_1$ are determined by column $b-1$. The procedure of calculating and sorting the $(L_0+1)(L_1+1)$ probability values is called \texttt{CalcProbAndSort}, as can be seen in Line~\ref{line:TGRAND1_sort} of Algorithm~\ref{alg:TGRAND1}. This section investigates the computational complexity of \texttt{CalcProbAndSort}.

Recall from Section~\ref{subsec:focus_on_TGRAND} that the $(L_0+1)(L_1+1)$ probability values compose the entries of matrix $\mathbf{F}$. Let $f_0(\ell_0)=p_{01}^{\ell_0}(1-p_{01})^{L_0-\ell_0}$ and $f_1(\ell_1)=p_{10}^{\ell_1}(1-p_{10})^{L_1-\ell_1}$. Matrix $\mathbf{F}$ is a rank-one matrix, which is generated as a trace of function $f(\ell_0,\ell_1)=f_0(\ell_0)f_1(\ell_1)$ on the two-dimensional uniform tensor product grid $\{(\ell_0,\ell_1)\}$:
\begin{equation}
 \mathbf{F}
  =
   \begin{bmatrix}  f_0(0) \\  f_0(1) \\ \vdots \\  f_0(L_0) \end{bmatrix}
   \begin{bmatrix}  f_1(0) & f_1(1) & \cdots & f_1(L_1) \end{bmatrix}.
\end{equation}
When the transition probabilities take the values $p_{01}=0$, $p_{01}=1$, $p_{10}=0$ or $p_{10}=1$, the singular case of $\mathbf{F}=\mathbf{0}$ is observed. If $p_{01}=0$ or $p_{10}=0$, the channel model has one state only. If $p_{01}=1$ or $p_{10}=1$, the channel has oscillatory memory, that is, the probability of remaining in a state is lower than the steady-state probability of being in that state~\cite{Mushkin1989}. These extreme cases are not of interest in the context of this paper and are ignored. We assume that $0<p_{01}<1$ and $0<p_{10}<1$ to ensure that all entries of $\mathbf{F}$ are positive, i.e., $f(\ell_0,\ell_1)>0$.

Since $f(\ell_0,\ell_1)$ is a product of probabilities, it could take very small values that would lead to an arithmetic underflow. A well-known approach to this problem is to operate in the log-domain and convert the products of probabilities into sums of log-probabilities\footnote{Although any base can be used for the $\log$ function, numerical examples in this paper use the logarithm to the base $2$.}.
Noting that
\begin{alignat}{2}
\label{eq:logprob}
  -\log f(\ell_0,\ell_1)
   & =
     &&-\ell_0 \log(p_{01}) - (L_0-\ell_0) \log(1-p_{01})\nonumber
  \\ &&&-\ell_1 \log(p_{10}) - (L_1-\ell_1) \log(1-p_{10})\nonumber
  \\ & =
     &&\underbrace{\ell_0 \log\Bigl(\frac{1-p_{01}}{p_{01}}\Bigr)}_{\phi_0(\ell_0)}
   + \underbrace{\ell_1 \log\Bigl(\frac{1-p_{10}}{p_{10}}\Bigr)}_{\phi_1(\ell_1)}\nonumber
  \\ &&& - \underbrace{\Bigl(L_0\log(1-p_{01}) +  L_1\log(1-p_{10})\Bigr)}_{\mathrm{\xi}}\nonumber
  \\ & =
  &&\underbrace{\phi_0(\ell_0) + \phi_1(\ell_1)}_{\phi(\ell_0,\ell_1)} \;-\; \mathrm{\xi},
\end{alignat}
we consider the matrix
\(
\mathbf{\Phi}=\begin{bmatrix}\phi(\ell_0,\ell_1)\end{bmatrix}
 \)
 with elements
 \(
\phi(\ell_0,\ell_1) = \phi_0(\ell_0) + \phi_1(\ell_1)
 \)
and of the same size as $\mathbf{F}$. The constant term $\xi$ in \eqref{eq:logprob} contributes equally to all log-probability values and can be ignored. The negative logarithm is a monotonically decreasing function, hence the largest element of $\mathbf{F}$ corresponds to the smallest element of $\mathbf{\Phi}.$
Consequently, instead of arranging the elements of $\mathbf{F}$ in decreasing order, we can now sort the elements of $\mathbf{\Phi}$ in increasing order.

The values of
\begin{equation}\label{eq:alpha}
 \alpha_0 = \log\Bigl(\frac{1-p_{01}}{p_{01}}\Bigr)
 \quad\text{and}\quad
 \alpha_1 = \log\Bigr(\frac{1-p_{10}}{p_{10}}\Bigl)
\end{equation}
can be computed and stored prior to the execution of the transversal GRAND algorithm.
Calculation of $\phi_0(\ell_0) = \alpha_0\ell_0$ requires one multiplication.
Alternatively, $\phi_0(\ell_0)$ can be written in recursive form as
\(
\phi_0(\ell_0)=\phi_0(\ell_0-1)+\alpha_0
\)
and hence, calculation of $\phi_0(\ell_0)$ for $\ell_0=0,\ldots,L_0$ involves $L_0$ additions.
Using the same reasoning, we deduce that $L_1$ additions are required for the calculation of $\phi_1(\ell_1)$ for $\ell_1=0,\ldots,L_1$. Having obtained the $(L_0+1)$ values of $\phi_0(\ell_0)$ and the $(L_1+1)$ values of $\phi_1(\ell_1)$, the procedure continues with the evaluation of the $(L_0+1)(L_1+1)$ entries of matrix $\mathbf{\Phi}$.
One sum is evaluated for every entry of matrix $\mathbf{\Phi}$, according to \eqref{eq:logprob}.
Therefore, $(L_0+1)(L_1+1)+L_0+L_1$ additions are needed for the computation of matrix $\mathbf{\Phi}.$

After \texttt{CalcProbAndSort} computes the entries $\phi(\ell_0,\ell_1)$ and inserts them in matrix $\mathbf{\Phi}$, it is tasked with arranging them in increasing order. Available sorting algorithms achieve different tradeoffs between runtime, stability and memory usage. Low overall computational complexity requires a short runtime, which translates into a small number of executed comparisons. If $n$ is the number of elements to be sorted, the worst-case number of comparisons is $O(n\log n),$ e.g., using the Heapsort~\cite{Williams1964} algorithm. The fact that Heapsort is not a stable algorithm does not affect the sorting process, as the relative order of equal entries in matrix $\mathbf{\Phi}$ does not have to be preserved. For this reason, Heapsort is adopted by \texttt{CalcProbAndSort} to sort the $n=(L_0+1)(L_1+1)$ elements of matrix $\mathbf{\Phi}$.

Recall that the output of \texttt{CalcProbAndSort} is not a sequence of sorted probability values but a sequence of coordinates $\{(\ell^{(\ell)}_0,\ell^{(\ell)}_1)\}$, which point to the locations of the sorted probability values in matrix $\mathbf{\Phi}$. The following section examines the relationship between $\phi(\ell_0,\ell_1)$ and the structure of matrix $\mathbf{\Phi}$ in order to develop a procedure that generates the coordinates of the ordered probability values without the need to first calculate all of the probability values.


\section{Implementation of a Tracing Procedure}
\label{sec:Tracing}

This section introduces \texttt{TraceSortedProb}, a procedure that traces negative log-probability values, from the smallest to the largest, without first calculating them and then sorting them, and outputs their coordinates. \texttt{TraceSortedProb} is proposed as an efficient alternative to \texttt{CalcProbAndSort}. The structural properties of matrix $\mathbf{\Phi}$ that \texttt{TraceSortedProb} takes advantage of are described in detail and the complexity of the proposed procedure is contrasted with that of \texttt{CalcProbAndSort}.

\subsection{Fundamentals of the Tracing Procedure}


Our goal is to trace the smallest elements of 
\(
\mathbf{\Phi}\!=\!\begin{bmatrix}\phi(\ell_0,\ell_1)\end{bmatrix}
 \), taking in account that
 \(
\phi(\ell_0,\ell_1) = \phi_0(\ell_0) + \phi_1(\ell_1),
 \)
 where
\(
\phi_0(\ell_0)=\alpha_0\ell_0
\)
and
\(
\phi_1(\ell_1)=\alpha_1\ell_1.
\)
The value of $\phi(\ell_0,\ell_1)$ occupies the entry in row $\ell_0+1$ and column $\ell_1+1$ of matrix $\mathbf{\Phi}$. We first consider $\phi_0(\ell_0)$ and observe that, according to~\eqref{eq:alpha}, $\alpha_0>0$ for $p_{01}<\tfrac12$, $\alpha_0<0$ for $p_{01}>\tfrac12$, and $\alpha_0=0$ for $p_{01}=\tfrac12$. The sequence $\{\phi_0(\ell_0)\}$ for $\ell_0=0,\ldots,L_0$ is monotonically increasing for $\alpha_0>0$, monotonically decreasing for $\alpha_0<0$, and constant for $\alpha_0=0$. The same conclusions can be drawn for the sequence $\{\phi_1(\ell_1)\}$ for $\ell_1=0,\ldots,L_1$.

\subsubsection{Identifying the smallest entry of matrix $\mathbf{\Phi}$}
\label{sec:smallest}

We established that the smallest element in sequence $\{\phi_i(\ell_i)\}$, where $i\in\{0,1\}$, is $\phi_i(0)$ for $\alpha_i\geq 0$, and $\phi_i(L_i)$ for $\alpha_i\leq 0$. Depending on the values of $\alpha_0$ and $\alpha_1$, one of the corner entries of matrix $\mathbf{\Phi}$, which hold $\phi(0,0)$, $\phi(L_0,0)$, $\phi(0,L_1)$ and $\phi(L_0,L_1)$, will be the smallest entry of the matrix. Henceforth, we assume that $\alpha_0>0$ and $\alpha_1>0$. As a result, the sequences $\{\phi_0(\ell_0)\}$ and $\{\phi_1(\ell_1)\}$ are both monotonically increasing and, therefore, the smallest entry of matrix $\mathbf{\Phi}$ is in row $1$ and column $1$, where $\phi(0,0)$ resides. Consequently, $(\ell^{(1)}_0,\ell^{(1)}_1)=(0,0)$.

The cases where $\alpha_0\leq 0$ or $\alpha_1\leq 0$ differ only in the direction in which sequences $\{\phi_0(\ell_0)\}$ and $\{\phi_1(\ell_1)\}$ increase. For example, let us assume that $p_{01}<\tfrac12$ and $p_{10}>\tfrac12$, hence, $\alpha_0>0$ and $\alpha_1<0$. This means that $\{\phi_0(\ell_0)\}$ is a monotonically increasing sequence, but $\{\phi_1(\ell_1)\}$ is a monotonically decreasing sequence. If we set $\tilde{p}_{10}=1-p_{10}<\tfrac12$ or, equivalently, $\tilde{\alpha}_1=-\alpha_1>0$, we obtain the monotonically increasing sequence $\{\tilde{\phi}_1(\ell_1)\}$, which satisfies $\tilde{\phi}_1(\ell_1)=\phi_1(L_1-\ell_1)$. The negative log-probability function $\tilde{\phi}(\ell_0,\ell_1)=\phi_0(\ell_0)+\tilde{\phi}_1(\ell_1)$ is used to generate the entries of matrix $\tilde{\mathbf{\Phi}}$. If the coordinate pair of the $\ell$-th smallest entry of matrix $\tilde{\mathbf{\Phi}}$ is $(\ell^{(\ell)}_0,\tilde{\ell}^{(\ell)}_1)$, then the $\ell$-th smallest entry of the original matrix $\mathbf{\Phi}$ will simply be $(\ell^{(\ell)}_0,\ell^{(\ell)}_1)=(\ell^{(\ell)}_0,L_1-\tilde{\ell}^{(\ell)}_1)$. On the other hand, the case of $p_{10}=\tfrac12$ and, thus, $\alpha_1=0$ leads to a constant sequence $\{\phi_1(\ell_1)\}$. Ordering the entries of $\mathbf{\Phi}$ becomes a trivial problem, as it depends solely on the monotonicity of sequence $\{\phi_0(\ell_0)\}$, which is determined by the sign of $\alpha_0$. The arguments that were made for $\alpha_1<0$ and $\alpha_1=0$ can also be made for $\alpha_0<0$ and $\alpha_0=0$.

\subsubsection{Identifying the second smallest entry of matrix $\mathbf{\Phi}$}
\label{sec:second_smallest}

Given that $\phi(\ell_0,\ell_1)=\phi_0(\ell_0)+\phi_1(\ell_1)$ and that $\phi(0,0)$ holds the smallest entry of matrix $\mathbf{\Phi}$ for $\alpha_0>0$ and $\alpha_1>0$, the second smallest entry will be either $\phi(1,0)$ or $\phi(0,1)$. To compare these two candidates for the second smallest entry, we evaluate the following differences:
\begin{equation}
 \begin{split}
  \phi(1,0) - \phi(0,0)
   & = \phi_0(1) - \phi_0(0)
     = \alpha_0 (1-0)
     =  \alpha_0,
  \\
  \phi(0,1) - \phi(0,0)
   & = \phi_1(1) - \phi_1(0)
     = \alpha_1 (1-0)
     = \alpha_1.
 \end{split}
\end{equation}
Therefore, the problem boils down to comparing $\alpha_0$ with $\alpha_1$. If $\alpha_0<\alpha_1$, then $\phi(1,0)$ is the second smallest entry of matrix $\mathbf{\Phi}$, hence $(\ell^{(2)}_0,\ell^{(2)}_1)=(1,0)$; otherwise, $(\ell^{(2)}_0,\ell^{(2)}_1)=(0,1)$. The process of pinpointing the second smallest entry of matrix $\mathbf{\Phi}$ is illustrated in Fig.~\ref{fig:sort}(a). Note that the coordinates of the smallest and second smallest entries of $\mathbf{\Phi}$, namely $(\ell^{(1)}_0,\ell^{(1)}_1)$ and $(\ell^{(2)}_0,\ell^{(2)}_1)$, have been determined by the values of $\alpha_0$ and $\alpha_1$ only.

\subsubsection{Generalization for the remaining entries of matrix $\mathbf{\Phi}$}
\label{sec:proc_generalisation}

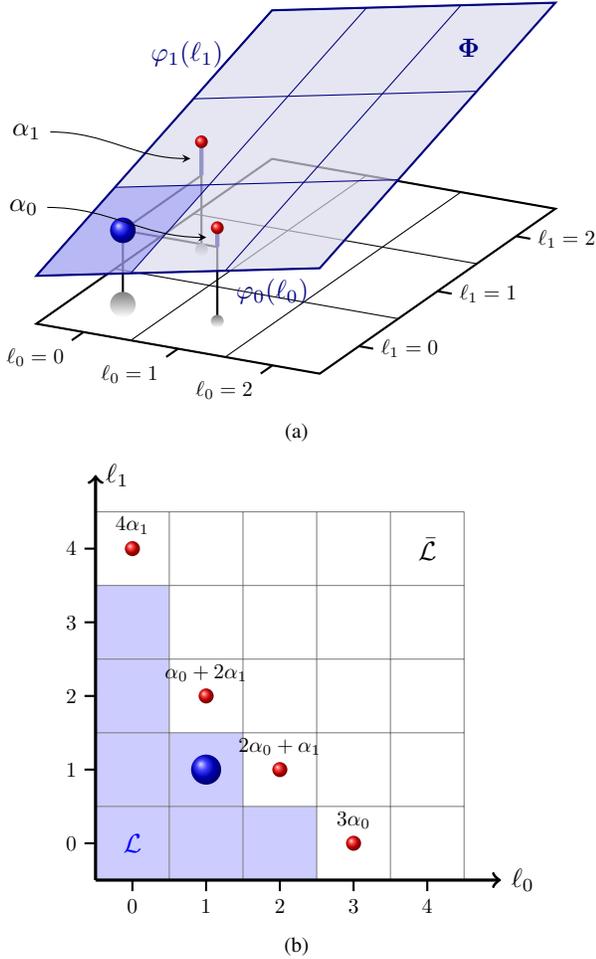
\begin{figure}[t]
 \begin{center}
\subfigure[]{
  \begin{tikzpicture}[scale=0.85, x={(-10:15mm)}, y={(35:15mm)}, z={(90:15mm)},
   declare function={z(\x,\y)=.5+.2*\x+.35*\y;}
   ]
   \def\nx{2} \def\ny{2}
   \def\fun{blue!50!black}
   \foreach \x in {0,1,2,...,\nx,\nx+1}{\draw[] (\x,0,0) -- (\x,\ny+1,0);}
   \foreach \y in {0,1,2,...,\ny,\ny+1}{\draw[] (0,\y,0) -- (\nx+1,\y,0);}
   \draw[thick] (0,0,0) -- (\nx+1,0,0) -- (\nx+1,\ny+1,0) -- (0,\ny+1,0) -- cycle;

   \foreach \x in {0,1,2,...,\nx}{\draw[thick] (\x+.5,0,0) --++ (0,-.15,0) node[below left,scale=.8]{$\ell_0=\x$} ;}
   \foreach \y in {0,1,2,...,\ny}{\draw[thick] (\nx+1,\y+.5,0) --++ (.15,0,0) node[right,scale=.8]{$\ell_1=\y$};}

   \foreach \x/\y in {0.5/0.5, 1.5/0.5, 0.5/1.5}{
    \draw[thick]  (\x,\y,0) --++(0,0,{z(\x,\y)});
   }
   \draw[thick] (0.5,0.5,{z(0.5,0.5)}) -- ++(1,0,0);
   \draw[thick] (0.5,0.5,{z(0.5,0.5)}) -- ++(0,1,0);
   \draw[\fun,ultra thick] (1.5,0.5,{z(1.5,0.5)}) -- node[midway](aim0){} (1.5,0.5,{z(0.5,0.5)});
   \draw[\fun,ultra thick] (0.5,1.5,{z(0.5,1.5)}) -- node[midway](aim1){} (0.5,1.5,{z(0.5,0.5)});

   \shade (0.5,0.5,0) circle(.2cm);
   \shade (1.5,0.5,0) circle(.1cm);
   \shade (0.5,1.5,0) circle(.1cm);

   \fill[white,opacity=.65] (0,0,{z(0,0)}) -- (\nx+1,0,{z(\nx+1,0)}) -- (\nx+1,\ny+1,{z(\nx+1,\ny+1)}) -- (0,\ny+1,{z(0,\ny+1)}) -- cycle;
   \fill[\fun,opacity=.1] (0,0,{z(0,0)}) -- (\nx+1,0,{z(\nx+1,0)}) -- (\nx+1,\ny+1,{z(\nx+1,\ny+1)}) -- (0,\ny+1,{z(0,\ny+1)}) -- cycle;
   \fill[blue,opacity=.2] (0,0,{z(0,0)}) -- (1,0,{z(1,0)}) -- (1,1,{z(1,1)}) -- (0,1,{z(0,1)}) -- cycle;

   \foreach \x in {0,1,2,...,\nx,\nx+1}{\draw[\fun] (\x,0,{z(\x,0)}) -- (\x,\ny+1,{z(\x,\ny+1)});}
   \foreach \y in {0,1,2,...,\ny,\ny+1}{\draw[\fun] (0,\y,{z(0,\y)}) -- (\nx+1,\y,{z(\nx+1,\y)});}
   \draw[\fun,thick] (0,0,{z(0,0)}) -- (\nx+1,0,{z(\nx+1,0)}) -- (\nx+1,\ny+1,{z(\nx+1,\ny+1)}) -- (0,\ny+1,{z(0,\ny+1)}) -- cycle;

   \shade[shading=ball,ball color=blue] (0.5,0.5,{z(0.5,0.5)}) circle(2mm);
   \shade[shading=ball,ball color=red] (1.5,0.5,{z(1.5,0.5)}) circle(1mm);
   \shade[shading=ball,ball color=red] (0.5,1.5,{z(0.5,1.5)}) circle(1mm);

   \node[\fun,left] at(0,{(\ny+1/2)},{z(0,(\ny+1/2))}){$\phi_1(\ell_1)$};
   \node[\fun,below] at({(\nx+1/2)},0,{z((\nx+1/2),0)}){$\phi_0(\ell_0)$};
   \node[\fun] at({\nx+1/2},{\ny+1/2},{z(\nx+1/2,\ny+1/2)}) {$\mathbf{\Phi}$};
   \draw[stealth-] (aim0) to[in=0,out=180] ++(-1.8,0,0)node[left]{$\alpha_0$};
   \draw[stealth-] (aim1) to[in=0,out=180] ++(-1.6,0,0)node[left]{$\alpha_1$};
  \end{tikzpicture}
  }
  \subfigure[]{
 \begin{tikzpicture}[scale=0.98, x=10mm,y=10mm]
   \def\nx{4} \def\ny{4}
   \fill[blue,opacity=.2] (0,0)rectangle(1,4)  (1,0)rectangle(2,2) (2,0)rectangle(3,1);
   \draw[gray] (0,0) grid ({\nx+1},{\ny+1});
   \draw[->,very thick] (0,0) -- ({\nx+1.5},0) node[right]{$\ell_0$};
   \draw[->,very thick] (0,0) -- (0,{\ny+1.5}) node[right]{$\ell_1$};
   \foreach \x in {0,1,2,...,\nx}{\draw[thick] (\x+.5,0) --++ (0,-.15) node[below,scale=.8]{$\x$} ;}
   \foreach \y in {0,1,2,...,\ny}{\draw[thick] (0,\y+.5) --++ (-.15,0) node[left,scale=.8]{$\y$} ;}
   \node(a)[circle,shading=ball,ball color=blue,minimum width=4mm,inner sep=0mm,outer sep=0ex] at(1.5,1.5){};
   \node(b)[circle,shading=ball,ball color=red,minimum width=2mm,inner sep=0mm,outer sep=0ex] at(0.5,4.5){};
   \node(c)[circle,shading=ball,ball color=red,minimum width=2mm,inner sep=0mm,outer sep=0ex] at(1.5,2.5){};
   \node(d)[circle,shading=ball,ball color=red,minimum width=2mm,inner sep=0mm,outer sep=0ex] at(2.5,1.5){};
   \node(e)[circle,shading=ball,ball color=red,minimum width=2mm,inner sep=0mm,outer sep=0ex] at(3.5,0.5){};
   \node[above,scale=.8]at(b.north){$4\alpha_1$};
   \node[above,scale=.8]at(c.north){$\alpha_0+2\alpha_1$};
   \node[above,scale=.8]at(d.north){$2\alpha_0+\alpha_1$};
   \node[above,scale=.8]at(e.north){$3\alpha_0$};
   \node[blue] at(.5,.5){$\L$};
   \node[black] at(4.5,4.5){$\bar\L$};
 \end{tikzpicture}
 }
 \end{center}
 \caption{Pictorial representation of how \texttt{TraceSortedProb} pinpoints the entries of matrix $\mathbf{\Phi}$, from the smallest to the largest, for $p_{01}<\tfrac12,$ \mbox{$p_{10}<\tfrac12$} and $p_{01} > p_{10}$. Equivalently, $\alpha_0>0$, $\alpha_1>0$ and $\alpha_0 < \alpha_1$. (a) A fragment of the matrix $\mathbf{\Phi}$ with the plane above visualizing the value of its elements. The smallest entry of $\mathbf{\Phi}$, which corresponds to the largest probability value, is at $(\ell_0^{(1)},\ell_1^{(1)})=(0,0)$ and is depicted by a large blue ball. Two candidates for the second smallest value are shown as small red balls at $(\ell_0,\ell_1)=(1,0)$ and $(\ell_0,\ell_1)=(0,1)$, where $\phi(\ell_0, \ell_1)$ takes the values $\alpha_0$ and $\alpha_1$, respectively. Since $\alpha_0 < \alpha_1$, the second smallest entry of $\mathbf{\Phi}$ is at $(\ell_0^{(2)},\ell_1^{(2)})=(1,0)$. (b) A larger fragment of matrix $\mathbf{\Phi}$. The shaded blue region represents the set $\L$ of the $\ell$ smallest entries of $\mathbf{\Phi}$. The largest element in $\L$ is depicted by a large blue ball. The candidates for the $(\ell+1)$-th smallest entry are the corners of $\bar\L$, shown as small red balls. Labels on candidate elements represent the values $\phi(\ell_0,\ell_1)=\alpha_0\ell_0+\alpha_1\ell_1$.
 }
 \label{fig:sort}
\end{figure}

To explain how \texttt{TraceSortedProb} proceeds after the two smallest entries of matrix $\mathbf{\Phi}$ have been identified, we start by making some simple observations that follow from the sum structure of $\phi(\ell_0,\ell_1)$ and the linearity of the univariate functions $\phi_0(\ell_0)$ and $\phi_1(\ell_1)$. Notice that the entries of matrix $\mathbf{\Phi}$ obey the following column-wise and row-wise properties:
  \begin{equation}\label{eq:lines}
   \begin{split}
    \ell_1'\leq\ell_1''\;\Leftrightarrow\; \phi(\ell_0,\ell_1') \leq  \phi(\ell_0,\ell_1'') &\;\, \text{for  $\ell_0=0,\dots,L_0,$}
    \\
    \ell_0'\leq\ell_0'' \;\Leftrightarrow\; \phi(\ell_0',\ell_1) \leq  \phi(\ell_0'',\ell_1) &\;\, \text{for  $\ell_1=0,\dots,L_1.$}
   \end{split}
  \end{equation}
To extend this observation, we introduce a weak partial order relation and a strong partial order relation for the coordinate pairs, denoted by $\preceq$ and $\prec$, respectively:
\begin{equation}\nonumber
 \begin{split}
  (\ell_0',\ell_1')\!\preceq\!(\ell_0'',\ell_1'')
    &\Leftrightarrow
    \ell_0'\leq\ell_0'' \,\text{and}\, \ell_1'\leq\ell_1'',
  \\
  (\ell_0',\ell_1')\!\prec\!(\ell_0'',\ell_1'')
    &\Leftrightarrow
    (\ell_0',\ell_1')\!\preceq\!(\ell_0'',\ell_1'')  \,\text{and}\,(\ell_0',\ell_1')\neq(\ell_0'',\ell_1'').
 \end{split}
\end{equation}
Then, it follows from~\eqref{eq:lines} that
\begin{equation}\label{eq:squares}
 (\ell_0',\ell_1')\preceq(\ell_0'',\ell_1'') \;\;\Rightarrow\;\; \phi(\ell_0',\ell_1') \leq  \phi(\ell_0'',\ell_1'').
\end{equation}

Suppose that the order of the $\ell$ smallest entries of matrix $\mathbf{\Phi}$ has been determined, that is:
\begin{equation}\label{eq:ordered_entries}
 \phi(\ell_0^{(1)},\ell_1^{(1)}) \leq
 \phi(\ell_0^{(2)},\ell_1^{(2)}) \leq \ldots \leq
 \phi(\ell_0^{(\ell)},\ell_1^{(\ell)}),
\end{equation}
and the $\ell$ coordinate pairs form the set \(
\L = \{(\ell_0^{(\ell')},\ell_1^{(\ell')})\}_{\ell'=1}^\ell
\), which has a particular geometric structure, as shown in Fig.~\ref{fig:sort}(b).
If
  $(\ell_0,\ell_1)\in\L$ and
  $(\ell_0',\ell_1')\preceq (\ell_0,\ell_1),$
  then from~\eqref{eq:squares} we have
  $\phi(\ell_0',\ell_1')\leq\phi(\ell_0,\ell_1),$
  thus $(\ell_0',\ell_1')\in\L.$ This means that every element $(\ell_0,\ell_1)\in\L$ defines a set of indices
\(
 \S(\ell_0,\ell_1) = \{(\ell_0',\ell_1'): (\ell_0',\ell_1') \preceq(\ell_0,\ell_1)\}
\) that make up a rectangle with coordinates $(0,0)$, $(0, \ell_1)$, $(\ell_0, \ell_1)$ and $(\ell_0, 0)$ in the grid specified by matrix $\mathbf{\Phi}$. The elements of  $\S(\ell_0,\ell_1)$ are also members of $\L$, i.e.,
\begin{equation}\label{eq:together}
 (\ell_0,\ell_1)\in\L
 \quad\Rightarrow\quad
 \S(\ell_0,\ell_1)\subseteq\L.
\end{equation}
Hence,
\(
\L = \bigcup_{\ell'=1,\ldots,\ell} \S(\ell_0^{(\ell')},\ell_1^{(\ell')}).
\)
Many of these rectangles are fully contained in others, and can be omitted from the union of sets that compose $\L$ without changing the result.
Rectangles that cannot be omitted are those that enclose elements, denoted as $(c_0,c_1)\in\L$, that are not contained in any other rectangular set:
\begin{equation}\label{eq:corners}
            (c_0,c_1):
            \quad \nexists (\ell_0,\ell_1)\in\L: \:
            (c_0,c_1) \prec (\ell_0,\ell_1).    
\end{equation}
We call such elements the \textit{corners} of $\L$, and denote the set of all corners of $\L$ as $\C_{\L}$. For example, if $\L$ is defined as in Fig.~\ref{fig:sort}(b), the set of all corners of $\L$ is given by $\C_{\L}=\{(0,3),(1,1),(2,0)\}$. By tracking $\C_{\L}$, we maintain the minimal representation of the coordinate set $\L$ at each step of the procedure, as
\(
 \L
   =
   \bigcup_{(c_0,c_1)\in\C_{\L}} \S(c_0,c_1).
\)
Furthermore, the element that was added last to~$\L$, that is,  $(\ell_0^{(\ell)},\ell_1^{(\ell)})$, is always a corner of $\L.$
Indeed, if $(\ell_0^{(\ell)},\ell_1^{(\ell)})\notin\C_{\L}$, then $\exists \ell'<\ell$ such that $(\ell_0^{(\ell)},\ell_1^{(\ell)})\prec(\ell_0^{(\ell')},\ell_1^{(\ell')})$, which leads to $\phi(\ell_0^{(\ell)},\ell_1^{(\ell)}) <\phi(\ell_0^{(\ell')},\ell_1^{(\ell')})$ according to \eqref{eq:squares}. The latter inequality contradicts \eqref{eq:ordered_entries} for $\ell'<\ell$, therefore the original assumption $(\ell_0^{(\ell)},\ell_1^{(\ell)})\!\notin\!\C_{\L}$ must be false, hence $(\ell_0^{(\ell)},\ell_1^{(\ell)})\!\in\!\C_{\L}$ must be true.

Observe in Fig.~\ref{fig:sort}(b) that, as $\ell$ increases, more entries of $\mathbf{\Phi}$ are traced, their coordinate pairs are added to $\L$, and $\L$ grows in size. The $(L_0+1)(L_1+1)-\ell$ coordinate pairs of the entries of $\mathbf{\Phi}$ that have not been considered yet in the tracing process compose the complementary set $\bar\L=\{(\ell_0,\ell_1)\notin\L\}$, which has structural properties that mirror those of $\L$. Whereas $\L$ holds the coordinates of the $\ell$ smallest entries of $\mathbf{\Phi}$, $\bar\L$ consists of the coordinates of the $(L_0+1)(L_1+1)-\ell$ largest entries of $\mathbf{\Phi}$. Whereas the coordinate pair that corresponds to the largest of the $\ell$ smallest entries of $\mathbf{\Phi}$ is one of the corners of $\L$, the coordinate pair that corresponds to the smallest of the $(L_0+1)(L_1+1)-\ell$ largest entries of $\mathbf{\Phi}$ is one of the corners of $\bar\L$. Let the corners of $\L$ be arranged in  $\C_{\L} =\{(c_0^{(1)},c_1^{(1)}),\ldots,(c_0^{(C)},c_1^{(C)})\}$ such that $c_0^{(1)} < \ldots < c_0^{(C)}$ and $c_1^{(1)} > \ldots > c_1^{(C)}$, where $C = |\C_{\L}|$. The corners of $\bar\L$ can then be calculated from the corners of $\L$ as follows:
\begin{equation}\label{eq:compl_corners}
\begin{split}
 \C_{\bar\L} = \{
  &\underbrace{(0,c_1^{(1)}+1)}_{\text{omit if $c_1^{(1)}\geq L_1$}},
 (c_0^{(1)}+1,c_1^{(2)}+1), \ldots \\[1.5ex]
 &\ldots, (c_0^{(C-1)}+1,c_1^{(C)}+1),
 \underbrace{(c_0^{(C)}+1,0)}_{\text{omit if $c_0^{(C)}\geq L_0$}}
 \}.
\end{split}
\end{equation}
In Fig.~\ref{fig:sort}(b), we mentioned that $\C_{\L}=\{(0,3),(1,1),(2,0)\}$, therefore $\C_{\bar\L}=\{(0,4),(1,2),(2,1),(3,0)\}$ based on \eqref{eq:compl_corners}.

To determine the $(\ell+1)$-th smallest entry of $\mathbf{\Phi}$ and obtain the pair $(\ell_0^{(\ell+1)},\ell_1^{(\ell+1)})$, we need to find the smallest entry among the candidates $(\ell_0,\ell_1)\in\C_{\bar\L},$ i.e.,
\begin{equation}
\label{eq:min_penalty}
\begin{split}
 (\ell_0^{(\ell+1)},\ell_1^{(\ell+1)})
    &= \underset{(\ell_0,\ell_1)\in\C_{\bar\L}}{\arg\,\min}\,\phi(\ell_0,\ell_1) \\
    &= \underset{(\ell_0,\ell_1)\in\C_{\bar\L}}{\arg\,\min} \left(\alpha_0\ell_0+\alpha_1\ell_1\right).
\end{split}
\end{equation}
The bivariate function $\phi(\ell_0,\ell_1)=\alpha_0\ell_0+\alpha_1\ell_1$ can be seen as a `penalty' that is minimized by \texttt{TraceSortedProb}. Fig.~\ref{fig:sort}(b) provides a visualization of the penalty for each of the four elements of $\C_{\bar\L}$.

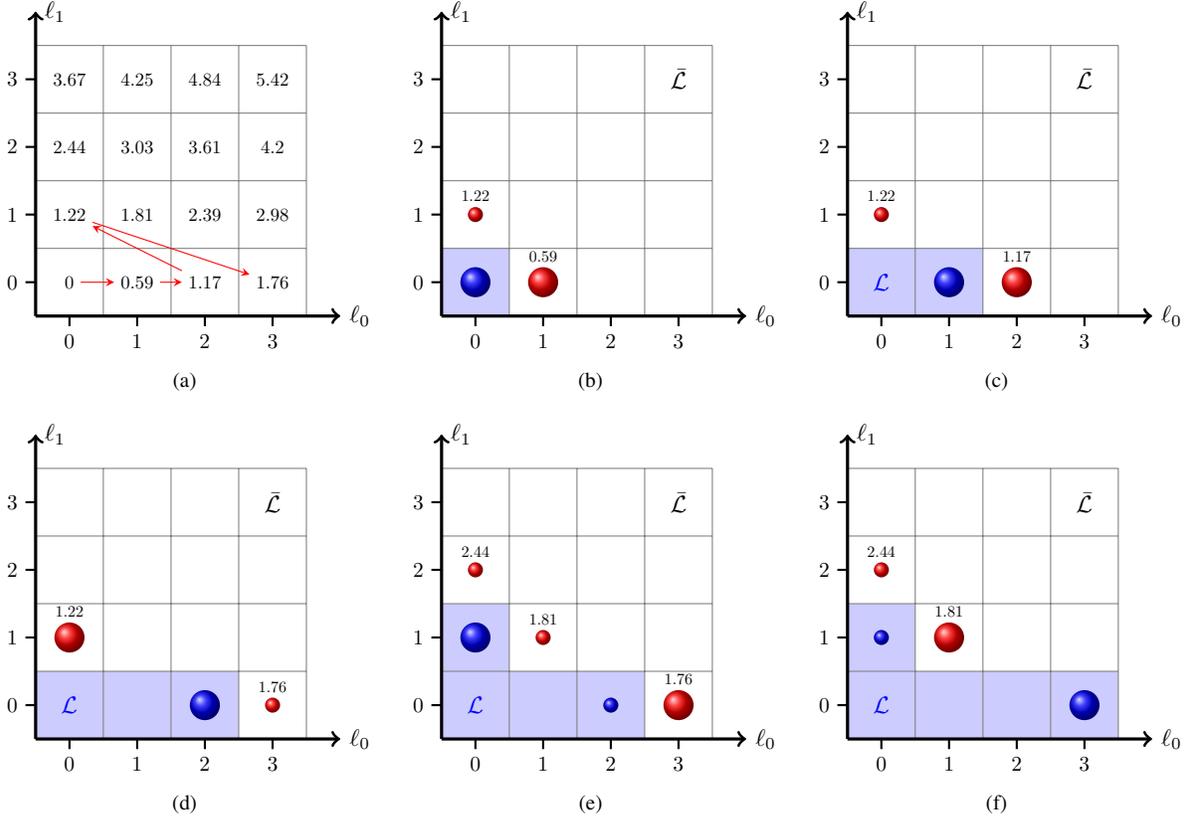
\begin{figure*}[t]
 \begin{center}
   \def\nx{3} \def\ny{3}
\subfigure[]{
 \begin{tikzpicture}[x=9mm, y=9mm]
   \draw[gray] (0,0) grid[step=1] ({\nx+1},{\ny+1});
   \draw[->,very thick] (0,0) -- ({\nx+1.5},0) node[right,scale=.9]{$\ell_0$};
   \draw[->,very thick] (0,0) -- (0,{\ny+1.5}) node[right,scale=.9]{$\ell_1$};
   \foreach \x in {0,1,2,...,\nx}{\draw[thick] (\x+.5,0) --++ (0,-.15) node[below,scale=.8]{$\x$} ;}
   \foreach \y in {0,1,2,...,\ny}{\draw[thick] (0,\y+.5) --++ (-.15,0) node[left,scale=.8]{$\y$} ;}
   \pgfmathsetmacro\pzero{0.4}
   \pgfmathsetmacro\pone{0.3}
   \pgfmathsetmacro\a{log2(1/\pzero -1)}
   \pgfmathsetmacro\b{log2(1/\pone -1)}
   \foreach \x in {0,1,2,...,\nx}{
    \foreach \y in {0,1,2,...,\ny}{
     \node(p\x\y)[scale=.7]at({\x+.5},{\y+.5}){$\pgfmathparse{\a*\x+\b*\y}\pgfmathprintnumber[precision=2]{\pgfmathresult}$};
    }
   }
   \draw[red,-stealth] (p00) to (p10);
   \draw[red,-stealth] (p10) to (p20);
   \draw[red,-stealth] (p20) to (p01);
   \draw[red,-stealth] (p01) to (p30);
 \end{tikzpicture}
  }
\subfigure[]{
 \begin{tikzpicture}[x=9mm,y=9mm]
   \fill[blue,opacity=.2] (0,0)rectangle(1,1);
   \draw[gray] (0,0) grid[step=1] ({\nx+1},{\ny+1});
   \draw[->,very thick] (0,0) -- ({\nx+1.5},0) node[right,scale=.9]{$\ell_0$};
   \draw[->,very thick] (0,0) -- (0,{\ny+1.5}) node[right,scale=.9]{$\ell_1$};
   \foreach \x in {0,1,2,...,\nx}{\draw[thick] (\x+.5,0) --++ (0,-.15) node[below,scale=.8]{$\x$} ;}
   \foreach \y in {0,1,2,...,\ny}{\draw[thick] (0,\y+.5) --++ (-.15,0) node[left,scale=.8]{$\y$} ;}
   \node(a)[circle,shading=ball,ball color=blue,minimum width=4mm,inner sep=0mm,outer sep=0ex] at(0.5,0.5){};
   \node(b)[circle,shading=ball,ball color=red,minimum width=2mm,inner sep=0mm,outer sep=0ex] at(0.5,1.5){};
   \node(c)[circle,shading=ball,ball color=red,minimum width=4mm,inner sep=0mm,outer sep=0ex] at(1.5,0.5){};
   \node[above,scale=.6]at(b.north){$1.22$};
   \node[above,scale=.6]at(c.north){$0.59$};
   \node[black,scale=.9] at(3.5,3.5){$\bar\L$};
 \end{tikzpicture}
 }
\subfigure[]{
 \begin{tikzpicture}[x=9mm,y=9mm]
   \fill[blue,opacity=.2] (0,0)rectangle(2,1);
   \draw[gray] (0,0) grid[step=1] ({\nx+1},{\ny+1});
   \draw[->,very thick] (0,0) -- ({\nx+1.5},0) node[right,scale=.9]{$\ell_0$};
   \draw[->,very thick] (0,0) -- (0,{\ny+1.5}) node[right,scale=.9]{$\ell_1$};
   \foreach \x in {0,1,2,...,\nx}{\draw[thick] (\x+.5,0) --++ (0,-.15) node[below,scale=.8]{$\x$} ;}
   \foreach \y in {0,1,2,...,\ny}{\draw[thick] (0,\y+.5) --++ (-.15,0) node[left,scale=.8]{$\y$} ;}
   \node(a)[circle,shading=ball,ball color=blue,minimum width=4mm,inner sep=0mm,outer sep=0ex] at(1.5,0.5){};
   \node(b)[circle,shading=ball,ball color=red,minimum width=2mm,inner sep=0mm,outer sep=0ex] at(0.5,1.5){};
   \node(c)[circle,shading=ball,ball color=red,minimum width=4mm,inner sep=0mm,outer sep=0ex] at(2.5,0.5){};
   \node[above,scale=.6]at(b.north){$1.22$};
   \node[above,scale=.6]at(c.north){$1.17$};
   \node[blue,scale=.9] at(.5,.5){$\L$};
   \node[black,scale=.9] at(3.5,3.5){$\bar\L$};
 \end{tikzpicture}
 }
 \\
 \subfigure[]{
 \begin{tikzpicture}[x=9mm,y=9mm]
   \fill[blue,opacity=.2] (0,0)rectangle(3,1);
   \draw[gray] (0,0) grid[step=1] ({\nx+1},{\ny+1});
   \draw[->,very thick] (0,0) -- ({\nx+1.5},0) node[right,scale=.9]{$\ell_0$};
   \draw[->,very thick] (0,0) -- (0,{\ny+1.5}) node[right,scale=.9]{$\ell_1$};
   \foreach \x in {0,1,2,...,\nx}{\draw[thick] (\x+.5,0) --++ (0,-.15) node[below,scale=.8]{$\x$} ;}
   \foreach \y in {0,1,2,...,\ny}{\draw[thick] (0,\y+.5) --++ (-.15,0) node[left,scale=.8]{$\y$} ;}
   \node(a)[circle,shading=ball,ball color=blue,minimum width=4mm,inner sep=0mm,outer sep=0ex] at(2.5,0.5){};
   \node(b)[circle,shading=ball,ball color=red,minimum width=4mm,inner sep=0mm,outer sep=0ex] at(0.5,1.5){};
   \node(c)[circle,shading=ball,ball color=red,minimum width=2mm,inner sep=0mm,outer sep=0ex] at(3.5,0.5){};
   \node[above,scale=.6]at(b.north){$1.22$};
   \node[above,scale=.6]at(c.north){$1.76$};
   \node[blue,scale=.9] at(.5,.5){$\L$};
   \node[black,scale=.9] at(3.5,3.5){$\bar\L$};
 \end{tikzpicture}
 }
  \subfigure[]{
 \begin{tikzpicture}[x=9mm,y=9mm]
   \fill[blue,opacity=.2] (0,0)rectangle(3,1) (0,1)rectangle(1,2);
   \draw[gray] (0,0) grid[step=1] ({\nx+1},{\ny+1});
   \draw[->,very thick] (0,0) -- ({\nx+1.5},0) node[right,scale=.9]{$\ell_0$};
   \draw[->,very thick] (0,0) -- (0,{\ny+1.5}) node[right,scale=.9]{$\ell_1$};
   \foreach \x in {0,1,2,...,\nx}{\draw[thick] (\x+.5,0) --++ (0,-.15) node[below,scale=.8]{$\x$} ;}
   \foreach \y in {0,1,2,...,\ny}{\draw[thick] (0,\y+.5) --++ (-.15,0) node[left,scale=.8]{$\y$} ;}
   \node[circle,shading=ball,ball color=blue,minimum width=2mm,inner sep=0mm,outer sep=.5ex] at(2.5,0.5){};
   \node(a)[circle,shading=ball,ball color=blue,minimum width=4mm,inner sep=0mm,outer sep=0ex] at(0.5,1.5){};
   \node(b)[circle,shading=ball,ball color=red,minimum width=4mm,inner sep=0mm,outer sep=0ex] at(3.5,0.5){};
   \node(c)[circle,shading=ball,ball color=red,minimum width=2mm,inner sep=0mm,outer sep=0ex] at(0.5,2.5){};
   \node(d)[circle,shading=ball,ball color=red,minimum width=2mm,inner sep=0mm,outer sep=0ex] at(1.5,1.5){};
   \node[above,scale=.6]at(b.north){$1.76$};
   \node[above,scale=.6]at(c.north){$2.44$};
   \node[above,scale=.6]at(d.north){$1.81$};
   \node[blue,scale=.9] at(.5,.5){$\L$};
   \node[black,scale=.9] at(3.5,3.5){$\bar\L$};
 \end{tikzpicture}
 }
   \subfigure[]{
 \begin{tikzpicture}[x=9mm,y=9mm]
   \fill[blue,opacity=.2] (0,0)rectangle(4,1) (0,1)rectangle(1,2);
   \draw[gray] (0,0) grid[step=1] ({\nx+1},{\ny+1});
   \draw[->,very thick] (0,0) -- ({\nx+1.5},0) node[right,scale=.9]{$\ell_0$};
   \draw[->,very thick] (0,0) -- (0,{\ny+1.5}) node[right,scale=.9]{$\ell_1$};
   \foreach \x in {0,1,2,...,\nx}{\draw[thick] (\x+.5,0) --++ (0,-.15) node[below,scale=.8]{$\x$} ;}
   \foreach \y in {0,1,2,...,\ny}{\draw[thick] (0,\y+.5) --++ (-.15,0) node[left,scale=.8]{$\y$} ;}
   \node[circle,shading=ball,ball color=blue,minimum width=2mm,inner sep=0mm,outer sep=.5ex] at(0.5,1.5){};
   \node(a)[circle,shading=ball,ball color=blue,minimum width=4mm,inner sep=0mm,outer sep=0ex] at(3.5,0.5){};
   \node(b)[circle,shading=ball,ball color=red,minimum width=2mm,inner sep=0mm,outer sep=0ex] at(0.5,2.5){};
   \node(c)[circle,shading=ball,ball color=red,minimum width=4mm,inner sep=0mm,outer sep=0ex] at(1.5,1.5){};
   \node[above,scale=.6]at(b.north){$2.44$};
   \node[above,scale=.6]at(c.north){$1.81$};
   \node[blue,scale=.9] at(.5,.5){$\L$};
   \node[black,scale=.9] at(3.5,3.5){$\bar\L$};
 \end{tikzpicture}
 }
 \end{center}
 \caption{Example demonstrating the operation of \texttt{TraceSortedProb} for $p_{01}=0.4$, $p_{10}=0.3$, $L_0=3$ and $L_1=3$.
 Based on these values, $\alpha_0=0.59$ and $\alpha_1=1.22$ (displayed values have been rounded up to two decimal places).
 The entries of $\mathbf{\Phi}$ as well as a path that begins at the smallest entry and ends at the fifth smallest entry of $\mathbf{\Phi}$ are presented in (a) for reference. The process of \texttt{TraceSortedProb} for identifying the~five smallest entries of $\mathbf{\Phi}$, without first calculating all entries and then ordering them, is described in (b)-(f). Blue balls of any size identify corners of~$\L$, while red balls of any size represent corners of $\bar\L$. A large blue ball marks the coordinate pair that has been added last to $\L$, as it points to the $\ell$-th smallest value of $\mathbf{\Phi}$, for $\ell=1$ in (b) to $\ell=5$ in (f).
 Labels on red balls represent the values of $\phi(\ell_0,\ell_1)$ at these points. A large red ball pinpoints the corner of $\bar\L$ that is chosen for a value of $\ell$ and changes to a large blue ball after it is added to $\L$ for the next value of $\ell$. Observe that the large blue balls in (b)-(f) occupy the same coordinates and in the same order as the coordinates of the five smallest entries~in~(a).
 }
 \label{fig:detailed_example}
\end{figure*}

The step-by-step process of \texttt{TraceSortedProb} to identify the five smallest entries of matrix $\mathbf{\Phi}$ for $\alpha_0=0.59$, $\alpha_1=1.22$, $L_0=3$ and $L_1=3$ is illustrated in Fig.~\ref{fig:detailed_example}. As explained in Section~\ref{sec:smallest} and shown in Fig.~\ref{fig:detailed_example}(b), the smallest entry is at $(0, 0)$ given that $a_0>0$ and $a_1>0$. Hence, $\C_\L=\{(0,0)\}$ for $\ell=1$, and $\C_{\bar\L}=\{(0,1), (1,0)\}$ based on \eqref{eq:compl_corners}. Using $\eqref{eq:min_penalty}$, we find that transitioning from $(0,0)$ to $(1,0)$ carries a lower penalty than transitioning to $(0,1)$. Therefore, the second smallest entry, i.e., for $\ell=2$, is at $(1,0)$. The pair $(1,0)$ is added to $\L$ and $\C_\L$, previous members of $\C_\L$ that do not comply with \eqref{eq:corners} are removed from $\C_\L$, the complementary set $\C_{\bar\L}$ is obtained from $\C_\L$ using \eqref{eq:compl_corners}, and the coordinate pair in $\C_{\bar\L}$ that minimizes $\eqref{eq:min_penalty}$ points to the location of the third smallest entry of $\mathbf{\Phi}$. This iterative process could continue until all entries of $\mathbf{\Phi}$ have been considered, but it could also terminate when the coordinates of a desired proportion of the smallest -- and, hence, most significant -- entries of $\mathbf{\Phi}$ have been determined.

\subsection{Complexity Analysis}
\label{sec:Comp_analysis}

As in \texttt{CalcProbAndSort}, the terms $\alpha_0=\log(\tfrac{1-p_{01}}{p_{01}})$ and $\alpha_1=\log(\tfrac{1-p_{10}}{p_{10}})$ are computed prior to the execution of the algorithm.
Part of the running time of \texttt{TraceSortedProb} is dedicated to the evaluation of the products $\alpha_0 \ell_0$ and $\alpha_1 \ell_1$ in~\eqref{eq:min_penalty} for all values of $\ell_0$ and $\ell_1$. Notice that as $\ell_0$ increases in steps of $1$, the value of $\alpha_0 \ell_0$ increases in steps of $\alpha_0$ (i.e., one addition per step). Calculation of the products $\alpha_0 \ell_0$ and $\alpha_1 \ell_1$, for $\ell_0=0,\ldots,L_0$ and $\ell_1=0,\ldots,L_1$, involves $L_0$ increments of $\ell_0$ and $L_1$ increments of $\ell_1$, which result in $L_0+L_1$ sequential additions. An extra addition is required to evaluate the penalty $\alpha_0 \ell_0 + \alpha_1 \ell_1$ whenever a new coordinate pair $(\ell_0,\ell_1)$ is inserted in $\C_{\bar\L}$.

As explained in Section \ref{sec:proc_generalisation}, \texttt{TraceSortedProb} maintains and updates $\C_\L$ and $\C_{\bar\L}$, which contain the corners of $\L$ and $\bar\L$, respectively. As the procedure traverses through the entries of matrix $\mathbf{\Phi}$ and the size of $\L$ increases in each step, maintaining and updating $\C_\L$ and $\C_{\bar\L}$ requires no more than $|\C_{\L}|$ operations per step. These operations between integer-valued indices can be omitted from the analysis, given that computational complexity is primarily driven by floating-point operations. At each step, the complexity of \texttt{TraceSortedProb} is dominated by the process of finding the lowest penalty among $|\C_{\bar\L}|$ penalties associated with the $|\C_{\bar\L}|$ corners of $\C_{\bar\L}$, as described by \eqref{eq:min_penalty} and depicted in Fig.~\ref{fig:detailed_example}. Let the $|\C_{\bar\L}|$ penalties be stored in the form of a linked list data structure, the elements of which are organized in ascending order. Therefore, the first element of the list always corresponds to the lowest penalty. With the aid of Fig.~\ref{fig:detailed_example}, one can observe that, at east step, a new coordinate pair -- on average -- is inserted in $\C_{\bar\L}$ and the corresponding penalty is computed and inserted in the list. Identification of the position in the sorted linked list where the new penalty value should be inserted requires $O(\log|\C_{\bar\L}|)$ comparisons. The lowest penalty value is subsequently removed from the linked list and the corresponding corner in $\C_{\bar\L}$ is moved to $\C_\L$. After $(L_0+1)(L_1+1)$ steps, when all entries of matrix $\mathbf{\Phi}$ have been considered, \texttt{TraceSortedProb} has computed and inserted $(L_0+1)(L_1+1)$ penalty values in the linked list, raising the total number of additions to $(L_0+1)(L_1+1)+L_0+L_1$ and the total number of comparisons to \mbox{$O((L_0\!+\!1)(L_1\!+\!1)\log|\C_{\bar\L}|)$}. As can be inferred from \eqref{eq:compl_corners}, the structure of $\L$ ensures that $|\C_{\bar\L}|\leq\min(L_0+1, L_1+1)$. The number of comparisons performed by \texttt{TraceSortedProb} is thus $O((L_0+1)(L_1+1)\log\min(L_0+1, L_1+1))$.

We established that both \texttt{CalcProbAndSort} and \texttt{TraceSortedProb} carry out \mbox{$(L_0+1)(L_1+1)+L_0+L_1$} additions. If we set $n=(L_0+1)(L_1+1)$ and \mbox{$L_0=L_1$} to simplify expressions, \texttt{CalcProbAndSort} performs $O(n\log n)$ comparisons, whereas \texttt{TraceSortedProb} performs $O(n\log(\sqrt{n}-1))$ comparisons. If we take into account that $n\log(\sqrt{n}-1)<\tfrac{n}{2}\log n$, we conclude that the worst-case number of comparisons in both procedures is $O(n\log n)$.

Although both procedures have computational complexities of the same order, the quick convergence of GRAND-based methods to optimal solutions \cite{Duffy2019} gives \texttt{TraceSortedProb} an edge on \texttt{CalcProbAndSort}. The error vector that is selected to be a column of matrix $\hat{\mathbf{E}}_{\overbar{\mathcal{R}}}$ is found among error vectors that have been born out of the first few elements of the output sequence of coordinate pairs. Assume that transversal GRAND introduces a threshold~$\ell_\mathrm{th}$, which limits the number of coordinate pairs that are considered. \texttt{CalcProbAndSort} still has to sort the $(L_0+1)(L_1+1)$ entries of matrix $\mathbf{\Phi}$ and then output the coordinates of only the $\ell_\mathrm{th}$ smallest entries. Therefore, the complexity of \texttt{CalcProbAndSort} remains constant and independent of $\ell_\mathrm{th}$. In contrast, \texttt{TraceSortedProb} will generate the sequence of $\ell_\mathrm{th}$ coordinate pairs after the first $\ell_\mathrm{th}$ steps and will then terminate execution, thus ignoring the remaining $(L_0+1)(L_1+1)-\ell_\mathrm{th}$ entries of matrix $\mathbf{\Phi}$. This means that the product $(L_0+1)(L_1+1)$ can be replaced by $\ell_\mathrm{th}$ in the complexity expressions derived in this section. A summary of the total number of operations performed by \texttt{CalcProbAndSort} and \texttt{TraceSortedProb} for the evaluation of a single column of $\hat{\mathbf{E}}_{\overbar{\mathcal{R}}}$ is provided in Table~\ref{tb:num_operations}. Note that for $n=(L_0+1)(L_1+1)$ and $L_0=L_1$, the number of additions carried out by \texttt{TraceSortedProb} reduces from $O(n)$ to $O(\sqrt{n})$, while the worst-case number of comparisons drops from $O(n\log n)$ to $O(\log n)$, provided that $\ell_\mathrm{th}$ is independent of $n$.

\begin{table}[!t]
\renewcommand{\arraystretch}{1.6}
\caption{Number of operations required by two competing procedures for the evaluation of a single column of the $L\times B$ matrix $\hat{\mathbf{E}}_{\overbar{\mathcal{R}}}$, where $L=L_0+L_1$.}%
\centering
\begin{tabular}{|l||c|}
\hline Operations      & \texttt{CalcProbAndSort} \\ \hline
\hline Additions & $(L_0+1)(L_1+1)+L_0+L_1$  \\
\hline \makecell[l]{Worst-case number\\of comparisons}  & \makecell[c]{$O(n\log n)$\\for $n=(L_0+1)(L_1+1)$}  \\
\hline
\end{tabular}

\bigskip
\noindent
\begin{tabular}{|l||c|}
\hline Operations      & \texttt{TraceSortedProb} \\ \hline
\hline Additions   & \makecell[c]{$\ell_\mathrm{th}+L_0+L_1$\\for $1\leq\ell_\mathrm{th}\leq(L_0+1)(L_1+1)$} \\
\hline \makecell[l]{Worst-case number\\of comparisons}  & \makecell[c]{$O(\ell_\mathrm{th}\log\min(L_0+1,L_1+1))$\\for $1\leq\ell_\mathrm{th}\leq(L_0+1)(L_1+1)$} \\
\hline
\end{tabular}
\label{tb:num_operations}
\end{table}


\section{Complexity Characterization based on System Parameters}
\label{sec:Comp_char}

The analysis in Sections~\ref{sec:Sorting} and \ref{sec:Comp_analysis} used the number of zeros and ones, denoted by $L_0$ and $L_1$ respectively, in column $b-1$ of the $L\times B$ matrix $\hat{\mathbf{E}}_{\overbar{\mathcal{R}}}$ to characterize the computational complexity associated with the generation of ordered coordinate pairs $(\ell_0^{(\ell)},\ell_1^{(\ell)})$ for the estimation of column $b$ of $\hat{\mathbf{E}}_{\overbar{\mathcal{R}}}$, where $\ell=1,\dots,(L_0+1)(L_1+1)$ in \texttt{CalcProbAndSort}, and $\ell=1,\dots,\ell_\mathrm{th}$ in \texttt{TraceSortedProb}. This section expresses the complexity of each procedure for the estimation of the full matrix $\hat{\mathbf{E}}_{\overbar{\mathcal{R}}}$ in terms of primary system parameters. 

Both syndrome decoding and transversal GRAND rely on RLC decoding, which has complexity $O(K^3)$ if Gaussian elimination is employed \cite{Chen2019}. As explained in Section~\ref{sec:SD}, syndrome decoding considers candidate error vectors for each column of $\hat{\mathbf{E}}_{\overbar{\mathcal{R}}}$ in the same order always, regardless of the channel parameters. In transversal GRAND, the ordering of the candidate vectors is informed by the channel parameters. As a result, transversal GRAND incurs additional complexity, which is further discussed in this section, but has the potential to obtain better estimates of the error matrix and, thus, stands a higher chance of recovering the source message, as demonstrated in Section~\ref{sec:Results}. 

\subsection{Distribution of the elements of the error matrix}

Section~\ref{sec:Comp_analysis} established that the complexity of the ordering procedures used in transversal GRAND depends on the number of zeros and ones in each column of $\hat{\mathbf{E}}_{\overbar{\mathcal{R}}}$. Recall that matrix $\hat{\mathbf{E}}_{\overbar{\mathcal{R}}}$ has dimensions $L\times B$, where $L=N-N_\mathrm{R}$ is the number of received coded packets that contain errors. The worst-case scenario for transversal GRAND occurs when errors are spread out over all $N$ coded packets, in which case the dimensions of $\hat{\mathbf{E}}_{\overbar{\mathcal{R}}}$ are $N\times B$. Let $X_b$ be a discrete random variable that represents the number of zeros in column $b$ of the $N\times B$ matrix $\hat{\mathbf{E}}_{\overbar{\mathcal{R}}}$. In other words, $X_b$ represents the number of independent Markov chains that are in state $0$ in step $b$, for $b=1,\ldots,B$. The number of ones in column $b$ is given by $N-X_b$. 
We denote the expected value of $X_b$ by $\mathrm{E}(X_b).$ 
The following recurrence relation can be obtained from the Markov chain in Fig.~\ref{fig:channel_model}:
\begin{equation}
\label{eq:MC_recursive}
\begin{split}
\mathrm{E}(X_b) &= (N-\mathrm{E}(X_{b-1})) p_{10} + \mathrm{E}(X_{b-1}) (1-p_{01})\\
&= Np_{10} + \mathrm{E}(X_{b-1}) \left(1-p_{10}-p_{01}\right),
\end{split}
\end{equation}
which states that the expected number of zeros in column $b$ is the sum of the expected number of ones in column $b-1$ that changed to zeros with probability $p_{10}$ and the expected number of zeros in column $b-1$ that did not change value with probability $1-p_{01}$. As mentioned in Section~\ref{subsec:focus_on_TGRAND}, the $N$ Markov chains start from state $0$, therefore $\mathrm{E}(X_0)=X_0=N$.

Iteration of \eqref{eq:MC_recursive} for an increasing value of $b$ reveals that $\mathrm{E}(X_b)$ can be written via geometric series as follows:
\begin{equation}
\label{eq:Xb_long}
\begin{split}
\mathrm{E}(X_b) &= N\left(1-p_{01}\sum_{i=0}^{b-1}(1-p_{10}-p_{01})^{i}\right)\\
&= N\left(1-p_{01}\left(\frac{1-(1-p_{10}-p_{01})^b}{p_{10}+p_{01}}\right)\right)\\
 &= N\left(1-\varepsilon\left(1-\mu^b\right)\right),
\end{split}
\end{equation}
where both the bit error probability $\varepsilon=p_{01}/(p_{01}+p_{10})$ and the channel memory $\mu=1-p_{10}-p_{01}$ have been defined in Section~\ref{sec:Gilbert}. For increasing $b$, the contribution of $\mu$ to the value of $\mathrm{W}(X_b)$ diminishes as the Markov chains approach the steady state. At the steady state, \eqref{eq:Xb_long} reduces to:
\begin{equation}
\label{eq:Xb_short}
\mathrm{E}(X) = N(1-\varepsilon),
\end{equation}
which is independent of $b$. 
Although elements in adjacent columns of $\hat{\mathbf{E}}_{\overbar{\mathcal{R}}}$ remain correlated at the steady state, we infer from \eqref{eq:Xb_short} that the number of zeros in each column of $\hat{\mathbf{E}}_{\overbar{\mathcal{R}}}$ follows a binomial distribution with parameters $N$ and $1-\varepsilon$. Hence, the number of zeros in any column of $\hat{\mathbf{E}}_{\overbar{\mathcal{R}}}$ can be represented by a single random variable $X\sim \text{Binomial}(N, 1-\varepsilon)$ with $\mathrm{E}(X)$ given by \eqref{eq:Xb_short}.

Knowledge of the distribution of $X$ can be used to estimate the number of a particular type of operations, e.g., additions or comparisons, performed by each of the ordering procedures presented in Sections~\ref{sec:Sorting}~and~\ref{sec:Comp_analysis}.

\subsection{Computational complexity of additions} \label{subsec:compl_add} 

Let $g^\text{sort}_\text{add}(B, N, \varepsilon)$ denote the average number of additions required for $B$ runs of \texttt{CalcProbAndSort} to obtain the $B$ columns of $\hat{\mathbf{E}}_{\overbar{\mathcal{R}}}$. From Table~\ref{tb:num_operations}, for $L_0\leftarrow X$ and $L_1\leftarrow N-X$, we obtain:
\begin{equation}
\label{eq:g_add_sort_1}
\begin{split}
g^\text{sort}_\text{add}(B, N, \varepsilon)&=\mathrm{E}\!\left(\sum_{b=1}^{B}\left((X+1)(N-X+1)+N\right)\right)\\
&=B\left(N\mathrm{E}(X)-\mathrm{E}(X^2)+2N+1\right),
\end{split}
\end{equation}
where
\begin{equation}
\label{eq:mean_sq_binomial}
\mathrm{E}(X^2)=\text{Var}(X)+\left(\mathrm{E}(X)\right)^2.
\end{equation}
Since $X\sim \text{Binomial}(N, 1-\varepsilon)$, the variance of $X$ is given by $\text{Var}(X)=N\varepsilon(1-\varepsilon)$ and the mean of $X$ is given by \eqref{eq:Xb_short}. Substituting $\text{Var}(X)$ and $\mathrm{E}(X)$ into \eqref{eq:g_add_sort_1} results in:
\begin{equation}
\label{eq:g_add_sort_2}
\begin{split}
   g^\text{sort}_\text{add}(B, N, \varepsilon)
    & = B\bigl(N(N-1)\varepsilon(1-\varepsilon)+2N+1\bigr)
  \\  & \leq BN^2\varepsilon+2BN+B,
\end{split}
\end{equation}
which leads us to the conclusion that:
\begin{equation}
\label{eq:g_add_sort_bigO}
g^\text{sort}_\text{add}(B, N, \varepsilon) = O(BN^2),
\end{equation}
for a given $\varepsilon>0$ as $B,N\rightarrow\infty$. Positive values, which are not too close to zero, are considered for the bit error probability $\varepsilon$ because transversal GRAND is triggered only when RLC decoding cannot cope with the large number of erroneous received coded packets.


The average number of additions required for $B$ runs of \texttt{TraceSortedProb} can be obtained following a similar line of reasoning as before and the help of Table~\ref{tb:num_operations}:
\begin{equation}
\label{eq:g_add_trace}
\begin{split}
g^\text{trace}_\text{add}(B, N, \varepsilon)&=\mathrm{E}\left(\sum_{b=1}^{B}\left(\ell_\mathrm{th}+N\right)\right)\\
&=B\mathrm{E}(\ell_\mathrm{th}) + BN.
\end{split}
\end{equation}
As we show in Section~\ref{sec:Results}, $\ell_\mathrm{th}$ can be kept small, therefore
\begin{equation}
\label{eq:g_add_trace_bigO}
g^\text{trace}_\text{add}(B, N, \varepsilon) = O(BN).
\end{equation}

\subsection{Computational complexity of comparisons} \label{subsec:compl_comp} 

Let $g^\text{sort}_\text{comp}(B, N, \varepsilon)$ denote the average number of comparisons required for $B$ runs of \texttt{CalcProbAndSort} to obtain the $B$ columns of $\hat{\mathbf{E}}_{\overbar{\mathcal{R}}}$ and $g^\text{trace}_\text{comp}(B, N, \varepsilon)$ denote the number of comparisons required by \texttt{TraceSortedProb} for the same task.

The $O(n \log n)$ worst-case estimates in Table~\ref{tb:num_operations} are for a comparison sort algorithm, such as heapsort~\cite{Williams1964}. 
The average number of operations in heapsort is $n\log_2 n,$ i.e. the the leading term constant is one.
Hence, the average complexity of one \texttt{CalcProbAndSort} run is 
\begin{equation}
n\log_2n = (L_0+1)(L_1+1) \left(\log_2(L_0+1) + \log_2(L_1+1)\right),\nonumber
\end{equation}
while the complexity of one \texttt{TraceSortedProb} run is 
\begin{equation}
\begin{split}
\ell_\mathrm{th}\log_2\min(L_0+1,&L_1+1) \\
&\leq \tfrac12 \ell_\mathrm{th}\left(\log_2(L_0+1)+\log_2(L_1+1)\right)\nonumber,
\end{split}
\end{equation}
thus \texttt{CalcProbAndSort} is at least $2 (L_0+1) (L_1+1) / \ell_\mathrm{th}$ times more complex than \texttt{TraceSortedProb}.

As before, we plug $L_0\leftarrow X$ and $L_1\leftarrow N-X$, and average over the binomial random variable $X.$
Since $\log$ is a concave function, we can bound $\log_2(L_0+1)+\log_2(L_1+1)$ from above as follows:
\begin{equation}
\begin{split}
\log_2(L_0+1)+\log_2(L_1+1) &\leq 2\log_2(L_0+L_1+2)\\
&= 2\log_2(N+2).\nonumber
\end{split}
\end{equation}
Substituting this result in the estimates above, and proceeding in a similar way to ~\eqref{eq:g_add_sort_2}, we obtain:
\begin{align}
 g^\text{sort}_\text{comp}&(B, N, \varepsilon) 
       \leq 2 B \mathrm{E}\left( (X+1)(N-X+1) \right) \log_2(N+2)\nonumber
   \\ & = 2B\left(N\mathrm{E}(X)-\mathrm{E}(X^2)+N+1\right) \log_2(N+2)\nonumber
   \\ & \leq 2BN^2\log_2(N+2)\varepsilon + 2B(N+1)\log_2(N+2)\nonumber
   \\ & = O(BN^2\log N).\label{eq:g_comp_sort}
\end{align}
Following a similar line of reasoning for the number of comparisons in \texttt{TraceSortedProb}, we get: 
\begin{equation} \label{eq:g_comp_trace}
\begin{split}
g^\text{trace}_\text{comp}(B, N, \varepsilon)
 & \leq B \mathrm{E}\left(\ell_\mathrm{th} \right)\log_2(N+2).
\end{split}
\end{equation}
As we show in numerical experiments in Section~\ref{sec:Results}, $\ell_\mathrm{th}$ can be kept small, therefore the tracing procedure is asymptotically faster -- with respect to the number of comparisons -- than the sorting procedure.

\subsection{Relative complexity of additions and comparisons}

Additions are, on average, more expensive than comparisons. To explain this fact, consider a pair of integers 
$a=\overline{a_s\ldots a_1a_0} = \sum_{k=0}^s a_k 2^k$ 
and 
$b=\overline{b_s\ldots b_1b_0} = \sum_{k=0}^s b_k 2^k $, 
written in big--endian notation, with $s$ bits $a_k,b_k\in\mathbb{F}_2$ for $k=0,\cdots,s.$
Computing their sum $c=a+b$ in the same form $c=\overline{c_{s+1}c_s\ldots c_1c_0}$ would require $2s$ bit operations, $c_k=a_k+b_k+r_{k-1} \mod 2$ for $k=1,\ldots,s$ where $r_{k-1}$ is the carry-over from the previous bit.

In contrast, a comparison of $a$ and $b$ would require to look at the most significant bits $a_s$ and $b_s$ first. If  $a_s>b_s$, we can instantly conclude that $a>b$. If $a_s<b_s$, then $a<b$, and only if $a_s=b_s$ we need to consider the subsequent digits. In binary arithmetic, $a_s=b_s$ happens with probability $1/2$. Hence, the comparison of $a$ and $b$ requires one bit-operation with probability $1/2$, two bit-operations with probability $1/2^2$, and so on. Overall, the average complexity of a comparison for numbers of length $s$ bits is equal to:
\begin{equation}\label{eq:comparison}
    \sum_{k=1}^s \frac{k}{2^k} = 2 - \frac{2+s}{2^s} < 2,
\end{equation}
i.e., a comparison requires, on average, less than two bit-operations.

To summarise, comparisons are on average $s$ times faster than additions, where $s=32$ for single-precision computations and $s=64$ for double-precision computations.
We can therefore consider additions to dominate the complexity of the whole computation, and neglect the effect of comparisons. 


\section{Results and Discussion}
\label{sec:Results} 

In this section, the bit error probability, $\varepsilon$, and the average length of an error burst, $\Lambda$, are used as the channel parameters; they can be derived from the transition probabilities, $p_{10}$ and $p_{01}$, as explained in Section~\ref{sec:Gilbert}. Recall that if a destination node receives \mbox{$L=N-N_\mathrm{R}$} erroneous coded packets of $B$ bits, transversal GRAND attempts to repair them by estimating the $L\times B$ matrix $\hat{\mathbf{E}}_{\overbar{\mathcal{R}}}$ column by column. Estimation of a column of $\hat{\mathbf{E}}_{\overbar{\mathcal{R}}}$ relies on the entries of the previous column. If the previous column contains $L_0$ zeros and $L_1$ ones, where $L_0+L_1=L$, transversal GRAND classifies all $2^L$ possible vectors that are candidates for the next column of $\hat{\mathbf{E}}_{\overbar{\mathcal{R}}}$ into $(L_0+1)(L_1+1)$ groups according to their probability of occurrence, as explained in Section~\ref{subsec:focus_on_TGRAND}.
The \texttt{TraceSortedProb} procedure, introduced in Section~\ref{sec:Tracing}, provides transversal GRAND with the option to identify only the $\ell_\mathrm{th}$ groups that are most likely to occur, instead of all $(L_0+1)(L_1+1)$ groups, thus reducing computational complexity without necessarily degrading the estimation accuracy of the algorithm.


The authors of~\cite{Mohammadi2016} considered memoryless channels and demonstrated the performance gain achieved by syndrome decoding when it is combined with RLC decoding, in terms of the overall \textit{decoding probability}, that is, the probability of a destination node recovering the $K$ source packets when $N$ coded packets have been transmitted. Since the proposed transversal GRAND algorithm is an extension of syndrome decoding for burst error channels, the objective of the following section is to compare the decoding probabilities of stand-alone RLC decoding, RLC decoding combined with syndrome decoding (\mbox{RLC with SD}) and RLC decoding combined with transversal GRAND (RLC with \mbox{T-GRAND}) for various channel configurations. 

\subsection{Impact of T-GRAND on the decoding probability}
\label{sec:Results_DecProb}

\begin{figure}[t]
\centering
\includegraphics[width=0.95\columnwidth]{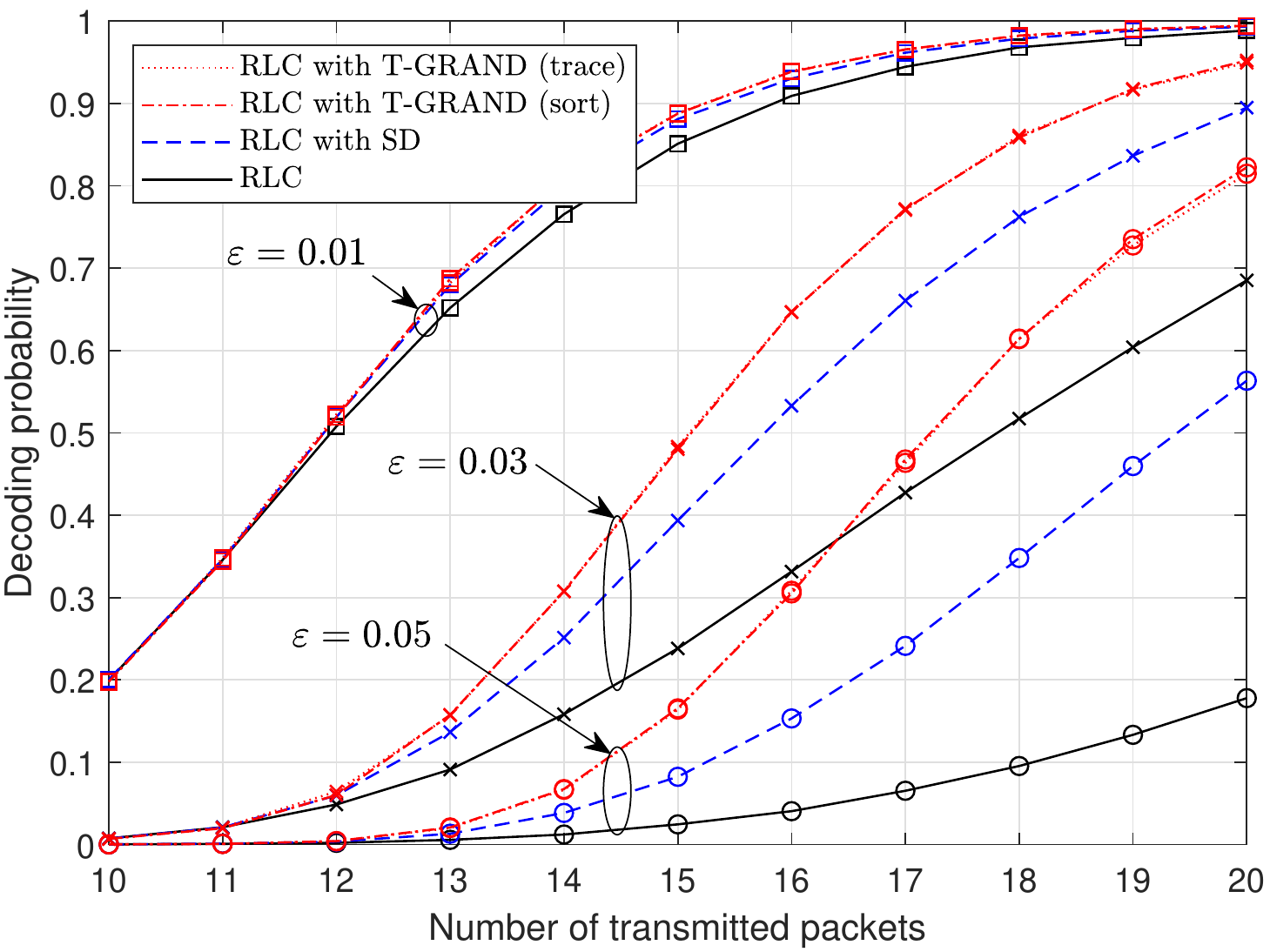}
\caption{Comparison of the decoding probabilities of RLC decoding, RLC decoding with SD, and RLC decoding with T-GRAND for $K=10$ source packets, $N$ transmitted coded packets, where $N=10,\ldots,20$, bit error probability $\varepsilon\in\{0.01,0.03,0.05\}$, average length of error bursts $\Lambda=4$ and packet length $B=64$.}
\label{fig:var_error_prob}
\end{figure}

\begin{figure}[t]
\centering
\includegraphics[width=0.95\columnwidth]{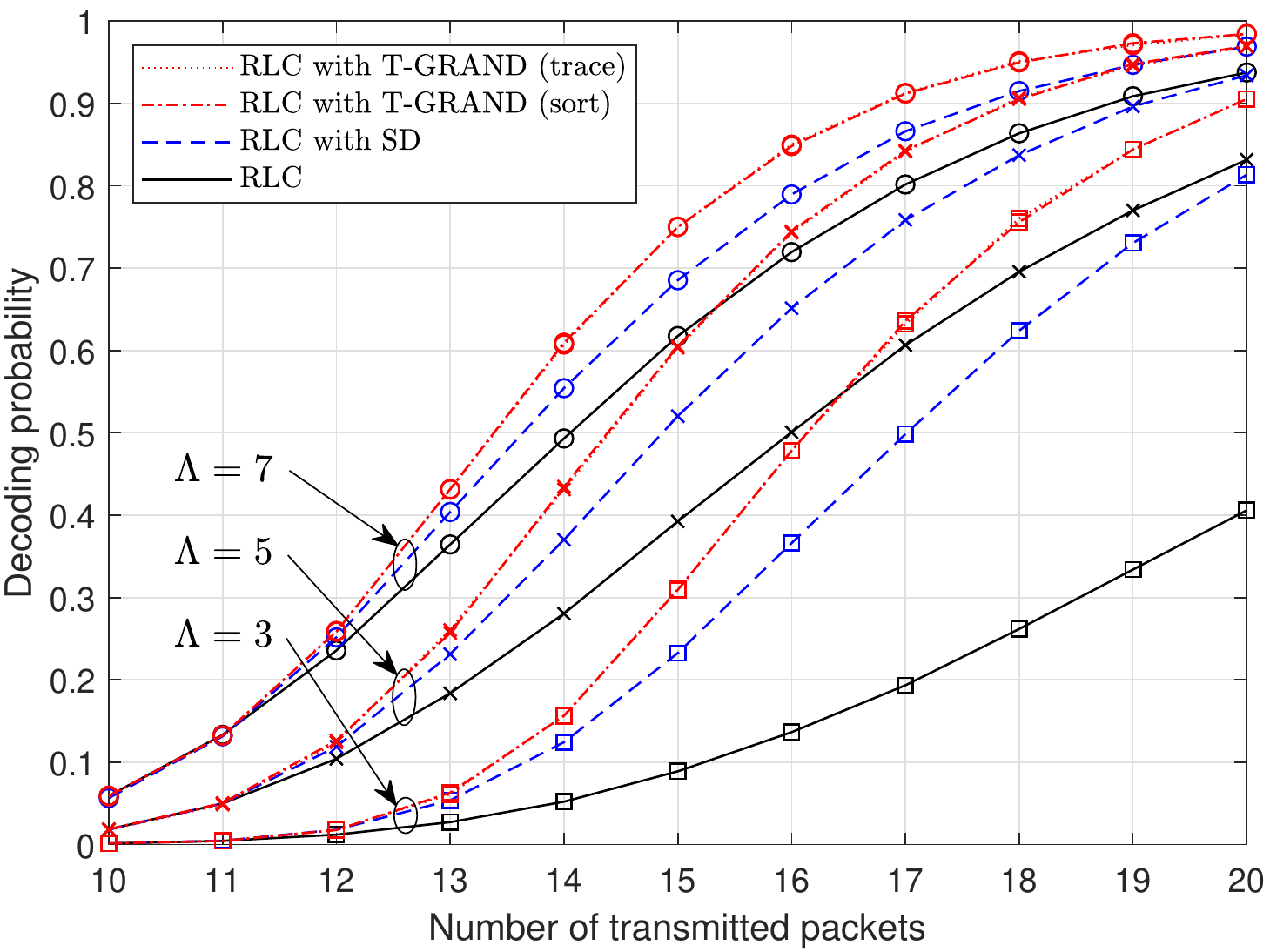}
\caption{Comparison of the decoding probabilities of RLC decoding, RLC decoding with SD, and RLC decoding with T-GRAND for $K=10$ source packets, $N$ transmitted coded packets, where $N=10,\ldots,20$, bit error probability $\varepsilon=0.03$, average length of error bursts $\Lambda\in\{3,5,7\}$ and packet length $B=64$.}
\label{fig:var_burst_length}
\end{figure}

\begin{figure}[t]
\centering
\includegraphics[width=0.95\columnwidth]{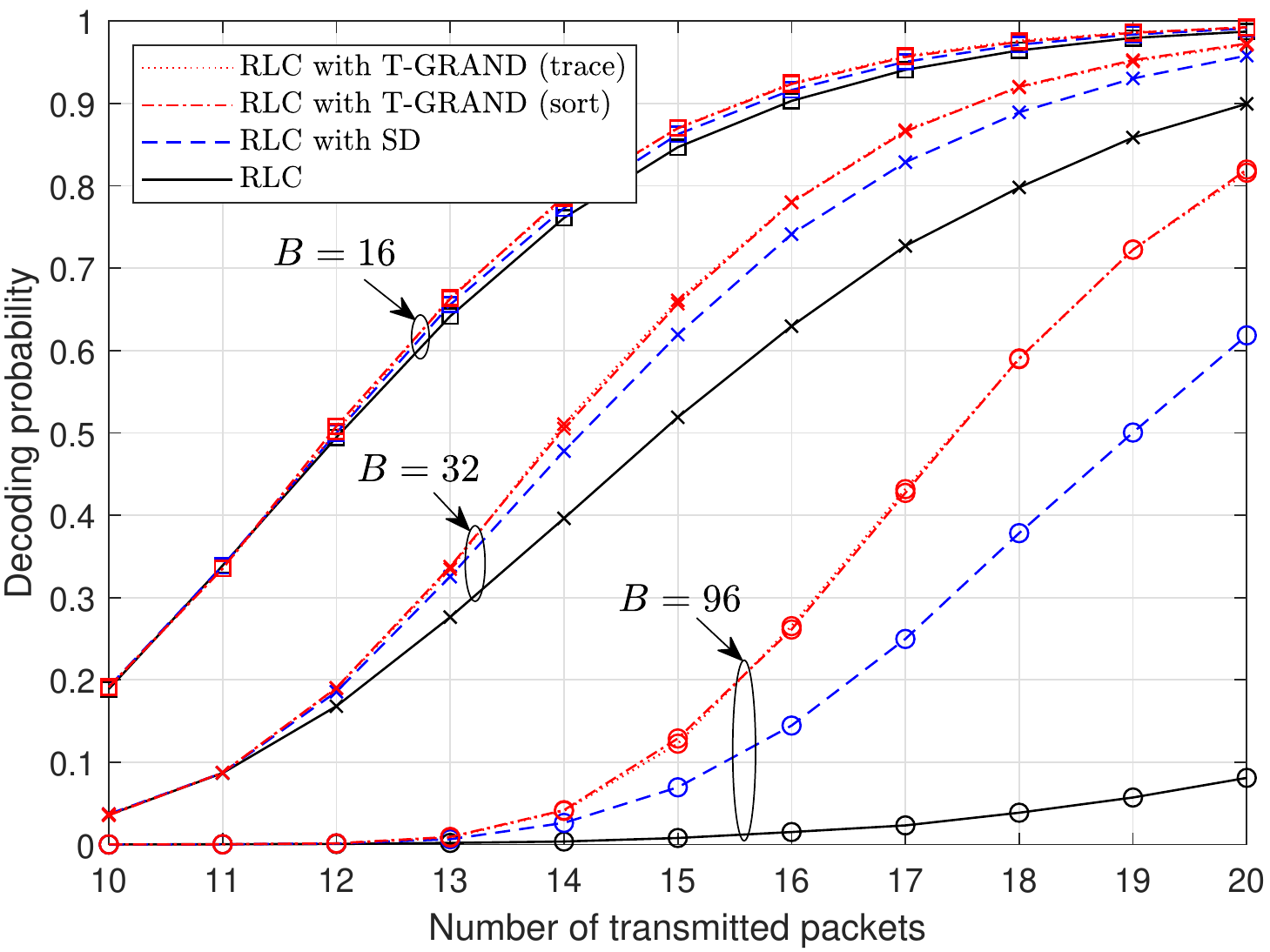}
\caption{Comparison of the decoding probabilities of RLC decoding, RLC decoding with SD, and RLC decoding with T-GRAND for $K=10$ source packets, $N$ transmitted coded packets, where $N=10,\ldots,20$, bit error probability $\varepsilon=0.03$, average length of error bursts $\Lambda=3$ and packet length $B\in\{16,32,96\}$.}
\label{fig:var_packet_length}
\end{figure}

The decoding probability was measured through simulations\footnote{Software simulations were implemented in MATLAB. The decoding probability shown in the plots for a tuple $(K, N, \varepsilon, \Lambda, B)$ has been averaged over $6\times10^{4}$ channel realizations. For each realization, the generator matrix of the RLC at the transmitter was randomly generated, as described in Section~\ref{sec:RLC}.} for $K=10$ source packets and a range of values for the number of transmitted packets ($N=10,11,\ldots,20$), the bit error probability ($\varepsilon\in\{0.01, 0.03, 0.05\}$), the average length of burst errors ($\Lambda\in\{3, 5, 7\}$), and the packet length in bits ($B\in\{16,32,96\}$). Measurements of the decoding probability are presented in Fig.~\ref{fig:var_error_prob}, Fig.~\ref{fig:var_burst_length} and Fig.~\ref{fig:var_packet_length}. In all three figures, T-GRAND implementations using \texttt{CalcProbAndSort} and \texttt{TraceSortedProb} have been considered. They have been abbreviated to ``T-GRAND (sort)'' and ``T-GRAND (trace)'', respectively. In the latter case, we set $\ell_\mathrm{th}=8$. The figures establish that the threshold $\ell_\mathrm{th}$ used by \texttt{TraceSortedProb} can be set to a small value to reduce the overall complexity of T-GRAND without compromising the decoding probability achieved by T-GRAND based on \texttt{CalcProbAndSort}.

Fig.~\ref{fig:var_error_prob} shows measurements for an increasing probability of the channel flipping bits in transmitted packets, when each packet contains $B=64$ bits and errors occur in bursts of length $\Lambda=4$ on average. For $\varepsilon=0.01$ and a growing number of transmitted packets, the destination node is likely to receive $K$ error-free linearly independent packets among the received packets and recover the $K$ source packets using RLC decoding; SD and T-GRAND offer only a marginal improvement in decoding probability. If the value of $\varepsilon$ is raised to $0.05$, the proportion of erroneous received packets increases, the RLC decoder experiences a notable drop in decoding performance, and SD and T-GRAND become instrumental in improving the chances of recovering the source packets. For example, when $\varepsilon=0.05$ and $N=20$, SD increases the decoding probability of RLC from $0.18$ to $0.56$ whereas T-GRAND, which considers correlations in errors when repairing coded packets, boosts the decoding probability to $0.82$.

Fig.~\ref{fig:var_burst_length} depicts the impact of the average length of error bursts on the decoding probability, when $\varepsilon=0.03$. When the value of $\varepsilon$ is constant, the same number of errors -- on average -- impair the transmitted packets, but errors aggregate in fewer packets as the average length of error bursts increases. Consequently, destination nodes receive an increasingly larger proportion of error-free packets as $\Lambda$ grows, and RLC decoding stands a greater chance of success. Although fewer packets are received in error for higher values of $\Lambda$, erroneous packets are more severely damaged by longer error bursts. Nevertheless, SD and especially T-GRAND can still improve the decoding probability of RLC. For example, the decoding probability of RLC, RLC with SD and RLC with T-GRAND is $0.72$, $0.79$ and $0.85$, respectively, for $\Lambda=7$ and $N=16$. On the other hand, low values of $\Lambda$ have a negative impact on RLC decoding; for a decreasing length of error bursts and a fixed average number of errors, the proportion of correctly received packets reduces. Packets corrupted by errors may dominate but errors are more sparsely distributed, and both SD and \mbox{T-GRAND} are more effective in repairing corrupted packets. Notice in Fig.~\ref{fig:var_burst_length} that the decoding probability of RLC, RLC with SD and RLC with T-GRAND is $0.41$, $0.81$ and $0.91$, respectively, for $\Lambda=3$ and $N=20$.

The effect of the packet length $B$ on the decoding probability is shown in Fig.~\ref{fig:var_packet_length}. For fixed values of $\varepsilon$ and $\Lambda$, the likelihood that a packet will be corrupted by errors increases with $B$ and, thus, the performance of RLC decoding deteriorates. When it comes to SD and T-GRAND, the value of $B$ also signifies the number of times that the algorithms have run in order to estimate the errors in the $B$ positions of the corrupted packets. The advantage of T-GRAND over SD becomes apparent as $B$ increases. For $B=96$ and $N=20$, SD lifts the decoding probability of RLC from $0.08$ to $0.62$; T-GRAND further increases the decoding probability to $0.82$ as is more successful than SD in making consecutive accurate guesses of error occurrences in the course of the $B=96$ runs.

\subsection{Impact of T-GRAND on the number of transmitted packets}
\label{sec:Results_TransmittedPackets}

\begin{figure}[t]
\centering
\includegraphics[width=0.95\columnwidth]{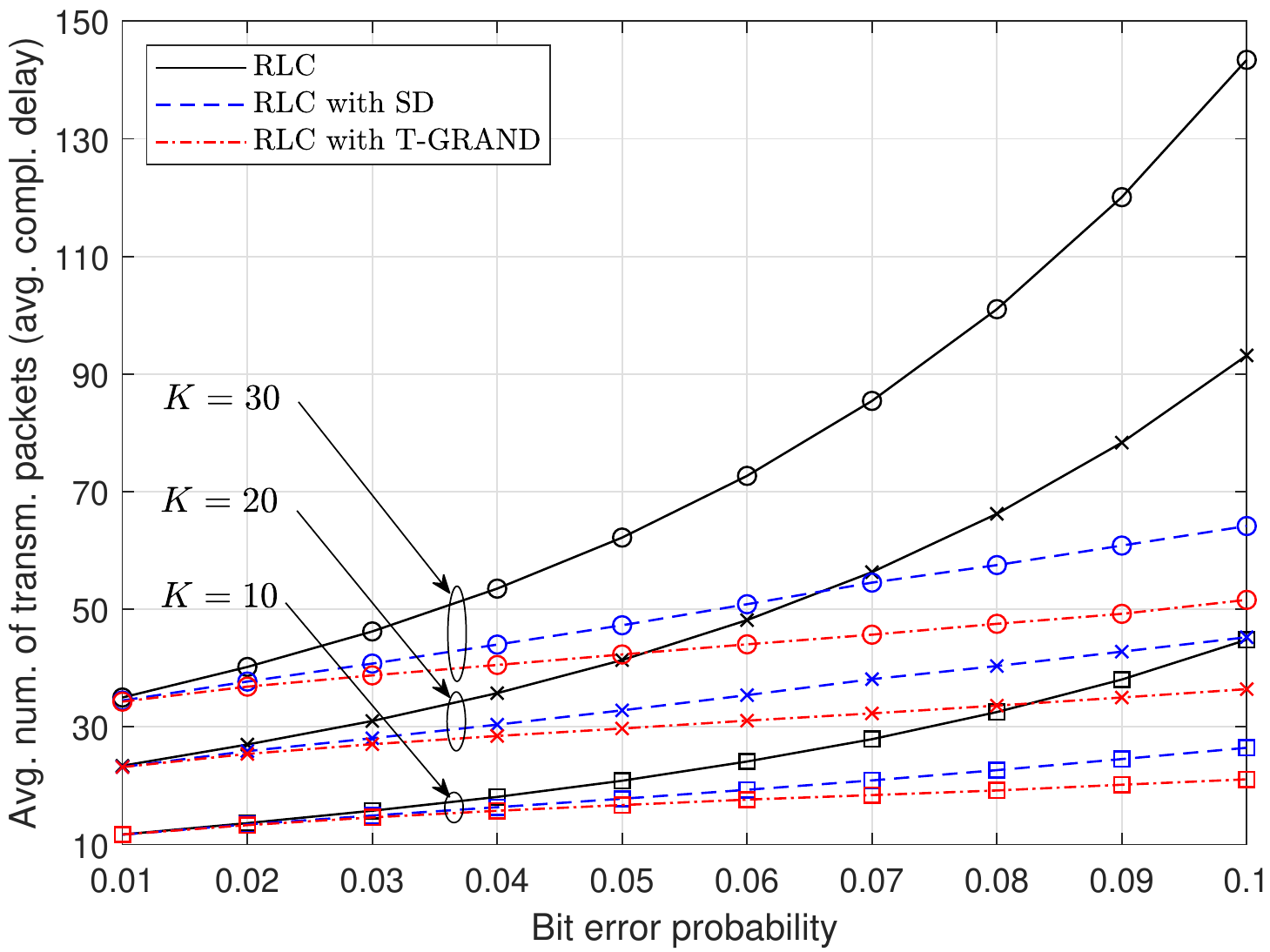}
\caption{Comparison of the average number of transmitted packets required by RLC decoding, RLC decoding with SD, and RLC decoding with T-GRAND to recover $K\in\{10,20,30\}$ source packets with decoding probability $100\%$ for a varying bit error probability $\varepsilon$. The average length of error bursts and the packet length have been set to $\Lambda=4$ and $B=64$, respectively.}
\label{fig:avgN_vs_epsilon}
\end{figure}

Whereas figures~\ref{fig:var_error_prob} to~\ref{fig:var_packet_length} depict the decoding probability as a function of the number of transmitted packets, Fig.~\ref{fig:avgN_vs_epsilon} shows the relationship between the \textit{average} number of transmitted packets and the bit error probability for a decoding probability equal to $1$. To obtain Fig.~\ref{fig:avgN_vs_epsilon}, simulations allowed the number of transmitted packets $N$ to increase until all $K$ source packets had been recovered by the destination node. The average number of packet transmissions for the successful delivery of the $K$ source packets is often referred to as the \textit{average completion delay}~\cite{Su2022}. In a broadcast scenario, a single destination node can be the major contributor to the average completion delay for the system, if it experiences the highest bit error probability compared to other destination nodes and, thus, frequently imposes the highest completion delay among all destination nodes, e.g., when it is located at the edge of the coverage area around the source node.

\begin{figure}[t]
\centering
\includegraphics[width=0.95\columnwidth]{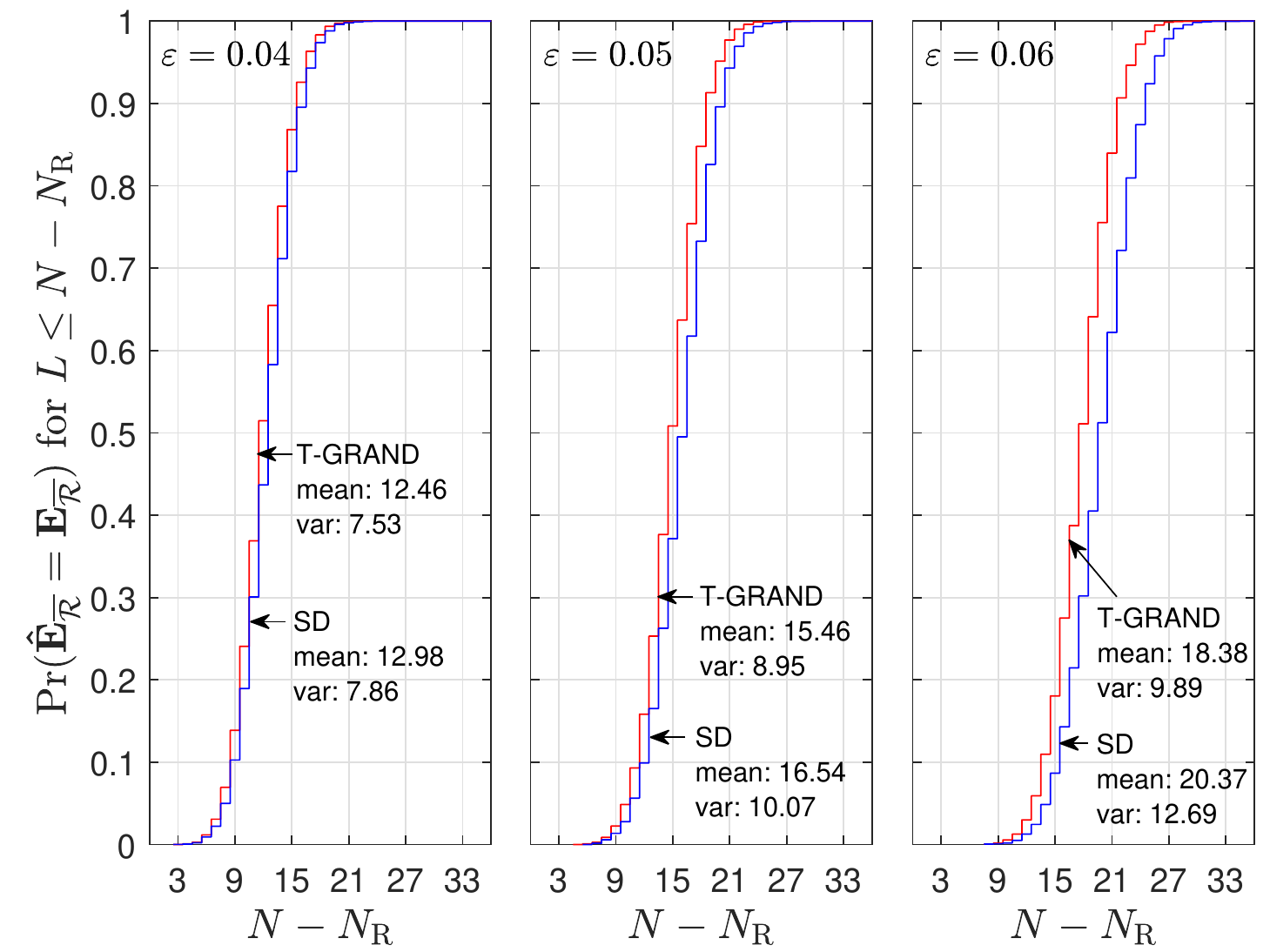}
\caption{Probability that the $L\times B$ estimated error matrix matches the actual error matrix when $L\leq N-N_\mathrm{R}$, where $N-N_\mathrm{R}$ is the number of packets received in error. The bit error probability $\varepsilon$ takes the values $0.04$ (left), $0.05$ (middle) and $0.06$ (right). The remaining parameters have been set as follows: $K=20$, $\Lambda=4$ and $B=64$.}
\label{fig:cumulative}
\end{figure}

If $\mathrm{E}(N)$ denotes the average number of transmitted packets, we observe that $\mathrm{E}(N)$ increases from $20.8$ (for $K=10$) to $41.3$ (for $K=20$) to $62.2$ (for $K=30$) in the case of RLC decoding, when $\varepsilon=0.05$. This is equivalent to $\mathrm{E}(N)\approx 2.07K$ for $K\in\{10,20,30\}$ and $\varepsilon=0.05$. Therefore, the ratio $\mathrm{E}(N)/K$ for a given bit error probability remains largely unaffected by $K$ when stand-alone RLC decoding is employed. When RLC decoding is combined with SD for $\varepsilon=0.05$, the ratio $\mathrm{E}(N)/K$ is smaller than $2.07$, and drops from $1.77$ (for $K=10$) to $1.57$ (for $K=30$). T-GRAND further reduces the requirement for a drastic increase in the value of $\mathrm{E}(N)$ to compensate for the increase in the bit error probability; when $\varepsilon=0.05$, the ratio $\mathrm{E}(N)/K$ decreases from $1.67$ (for $K=10$) to $1.41$ (for $K=30$). The gap in the average number of transmitted packets between T-GRAND and SD grows as the bit error probability increases. For instance, as can be seen in Fig.~\ref{fig:avgN_vs_epsilon}, $\mathrm{E}(N)=28.5$ for T-GRAND while $\mathrm{E}(N)=30.4$ for SD when $K=20$ and $\varepsilon=0.04$, i.e., \mbox{T-GRAND} requires $1.9$ fewer packets to be transmitted -- on average --  than SD to achieve a decoding probability of $1$. This difference increases to $3.1$ for $\varepsilon=0.05$, and to $4.3$ for $\varepsilon=0.06$. This example is used in Fig.~\ref{fig:cumulative} to further explore the potential of T-GRAND and SD to converge to a correct estimate of the error matrix.

Fig.~\ref{fig:cumulative} considers both T-GRAND and SD and depicts the probability that each method will correctly estimate the $L\times B$ matrix $\mathbf{E}_{\overbar{\mathcal{R}}}$ if $N-N_\mathrm{R}$ or fewer packets are received in error, i.e., $L\leq N-N_\mathrm{R}$, for $K=20$ and $\varepsilon\in\{0.04,0.05,0.06\}$. The three subfigures illustrate that T-GRAND can accurately estimate $\mathbf{E}_{\overbar{\mathcal{R}}}$ with a progressively higher chance than SD for a given number of erroneous received packets, as the value of $\varepsilon$ increases from $0.04$ to $0.06$. Fig.~\ref{fig:cumulative} also establishes that \mbox{T-GRAND} needs fewer received packets than SD to compute an accurate estimate of $\mathbf{E}_{\overbar{\mathcal{R}}}$ with the same probability as SD. Therefore, the results in Fig.~\ref{fig:cumulative} tally with the results in Fig.~\ref{fig:avgN_vs_epsilon}, which exhibited that T-GRAND requires fewer packet transmissions than SD to recover the source packets. 

\begin{figure}[t]
\centering
\includegraphics[width=0.95\columnwidth]{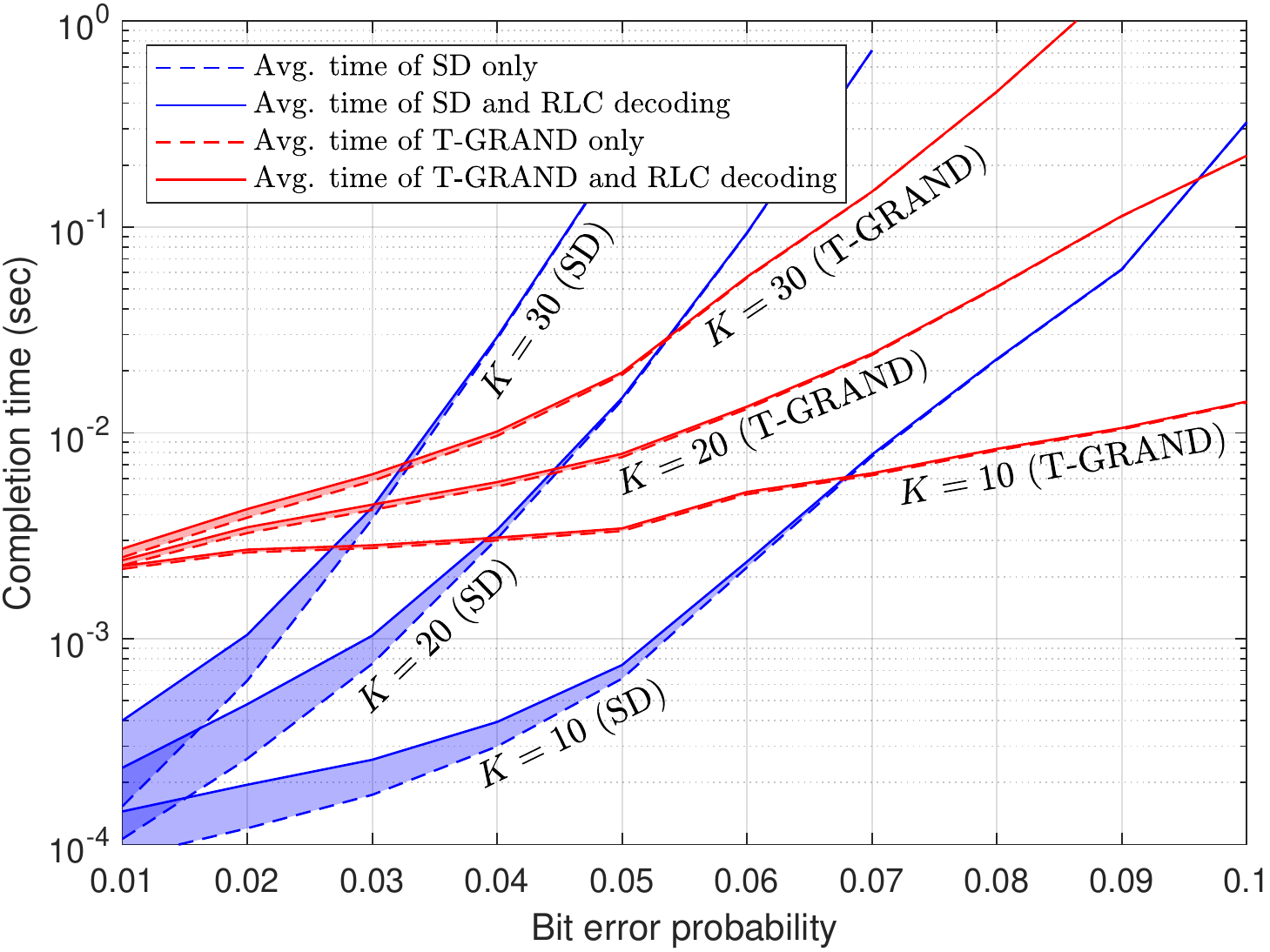}
\caption{Completion time of T-GRAND and SD, and overall completion time of a decoder that combines RLC decoding with T-GRAND or SD. Shaded regions represent the time added by RLC decoding to the overall completion time. The system configuration is identical to that considered in Fig.~\ref{fig:avgN_vs_epsilon}.}
\label{fig:time}
\end{figure}

\subsection{Impact of T-GRAND on the completion time}
\label{sec:Results_CompletionTime}

Recall that SD assumes that the channel is memoryless, even when it has memory, as is the case of this paper. As explained in Section~\ref{sec:SD}, the order with which SD generates and queries error vectors for the column-wise estimation of $\mathbf{E}_{\overbar{\mathcal{R}}}$ is fixed and independent of the channel parameters. \mbox{T-GRAND} incorporates an additional stage, which uses knowledge of the channel model to inform the ordering process, thus making the estimation of $\mathbf{E}_{\overbar{\mathcal{R}}}$ dependent on the channel parameters. This stage has given T-GRAND an advantage over SD in terms of the decoding probability (Figs.~\ref{fig:var_error_prob}-\ref{fig:var_packet_length}) and the average number of transmitted packets (Fig.~\ref{fig:avgN_vs_epsilon}). This section investigates the impact of this additional stage on the overall time complexity of T-GRAND and contrasts it to that of SD. The T-GRAND implementation that employs \texttt{CalcProbAndSort} has been considered to ascertain if a complexity gain over SD can be achieved, even when T-GRAND uses the more computational expensive of the two proposed ordering algorithms.

Fig.~\ref{fig:time} compares the average completion time\footnote{The simulations were carried out on a 64-bit Microsoft Windows 10 system using an Intel i7-1370P processor running at 1.9 GHz and 32 GB of RAM. Standard sequential programming in MATLAB was used for both SD and T-GRAND, that is, parallelization was not explored in this paper.} of the two error correction methods, that is, T-GRAND and SD. Solid lines represent the overall completion time, which includes the time taken by the initial attempt of RLC decoding to recover the source packets, the execution of the error correction method, and the subsequent successful RLC decoding of the error-free and repaired received packets. Dashed lines depict the average completion time of each error correction method, while shaded areas draw attention to the time added by the two RLC decoding operations to the overall completion time.

If Fig.~\ref{fig:time} is looked at in combination with Fig.~\ref{fig:avgN_vs_epsilon}, we observe that, for small values of $K$ and $\varepsilon$, SD and \mbox{T-GRAND} have a marginal difference in the average number of transmitted packets required to recover the source packets. This is because the pre-set ordering of SD and the adaptive ordering of T-GRAND generate similar outputs, although the adaptive capability of the latter method incurs a complexity cost. As the values of $K$ and $\varepsilon$ increase, adaptive ordering starts to exhibit a clearer advantage over pre-set ordering, as T-GRAND requires fewer packet transmissions than SD. Although this advantage of T-GRAND comes at the expense of added complexity, Fig.~\ref{fig:time} shows that the added computational cost is offset by the quicker convergence to a solution. Indeed, the pre-set ordering of SD results in the generation and querying of more error vectors than T-GRAND before an appropriate vector is identified for each column of $\hat{\mathbf{E}}_{\overbar{\mathcal{R}}}$. Therefore, for increasing values of $K$ and $\varepsilon$, T-GRAND has the potential to also achieve a shorter completion time than SD, in addition to the improved decoding probability and the reduced packet transmissions. 

\subsection{Discussion}
\label{sec:Discussion}

The results presented in this section established that \mbox{T-GRAND} has the potential to enhance the performance of an RLC decoder, that is, improve the probability of a destination node recovering the source packets, or reduce the required number of packet transmissions. The proliferation of RLC implementations, such as fountain coding and network coding, in vehicular networks was mentioned in Section~\ref{sec:intro}. RLC has been promoted as a way to combat intermittent connections, e.g., due to the high-speed movement of cars, or the sparse and highly dynamic traffic. For example, RLC has been proposed for comfort applications (e.g., multimedia streaming~\cite{Yousefi2010}), safety applications, (e.g., broadcast of post-crash warning messages~\cite{Abdullah2011}), and cooperative content distribution (e.g., sharing information about hazards~\cite{Ahmed2006}). In vehicle-to-vehicle scenarios, T-GRAND could improve the reliability of applications that rely on RLC decoding in adverse channel conditions. When RLC is used on vehicle-to-infrastructure links, T-GRAND could be integrated into the RLC decoder of roadside units to allow vehicles to transmit fewer packets without compromising reliability; therefore, T-GRAND could contribute to the improvement of the energy efficiency of vehicular communications and the reduction of the packet backlog at roadside units.



\section{Conclusions and Future Work}
\label{sec:Conclusions}

In this paper, we built on the `guessing random additive noise decoding' (GRAND) concept and developed transversal GRAND, a hard detection decoder that leverages correlations between channel errors and works in tandem with the decoder of a random linear code (RLC) at layers higher than the physical layer. Transversal GRAND generates and queries error vectors in order of likelihood, and attempts to correct bits that occupy the same position in erroneous received packets. After all bit positions have been considered, RLC decoding utilizes both repaired and correctly received packets to recover the original information packets. Calculation and sorting of the likelihood values of all error vectors for each bit position is a simple but computationally expensive procedure. For this reason, we also presented a computationally efficient method, which identifies the most likely error vectors without computing and ordering their likelihoods. Simulation results demonstrated that transversal GRAND has the potential to markedly improve the performance of RLC decoding when the channel introduces error bursts in transmitted packets.

The work presented in this paper has paved the way for new research directions, which we shall briefly review here. The first direction is the derivation of an expression for the probability of an RLC decoder recovering the source packets, when it is assisted by syndrome decoding or transversal GRAND. The decoding probability is directly related to the probability that the estimated error matrix is correct or, equivalently, each of the columns of the estimated error matrix has been identified correctly. The work by Galligan~\textit{et al.}~\cite{Galligan2023}, which was inspired by Forney's work on error exponents~\cite{Forney1968}, could be used to obtain an approximation of the likelihood that a column of the estimated error matrix is correct. The chain rule can then be applied to obtain an approximation of the joint probability that all columns of the error matrix have been estimated correctly. Of course, a product of approximated likelihoods will amplify the overall approximation error, but the product could still provide a useful bound on the probability that the estimated error matrix is accurate.

Another direction is the extension of transversal GRAND to RLCs over $\mathbb{F}_{2^q}=\{0,1,\ldots,2^q-1\}$ for $q\geq 1$. This would involve the replacement of the binary Gilbert-Elliott channel model by a non-binary model, e.g., \cite{Griffiths2008}, and the derivation of a multidimensional function that describes the probability of occurrence of each error vector, instead of the two-dimensional function given in \eqref{eq:prob_occurence}. The procedure for identifying the most likely error vectors would then have to be configured to trace a path of ordered negative log-probability values, not in a two-dimensional matrix as in Fig.~\ref{fig:detailed_example}, but in a multidimensional matrix. The generalization of transversal GRAND for RLCs over $\mathbb{F}_{2^q}$ will allow comparison of its performance with that of other non-binary schemes, e.g., PRAC~\cite{Angelopoulos2014}. 

A third, and equally interesting direction, is the investigation of transversal GRAND from an implementation perspective. Abbas~\textit{et al.}~\cite{Abbas2023} compiled a treatise on hardware architectures of GRAND variants for the physical layer. Although~\cite{Abbas2023} could inform aspects of the design of transversal GRAND, random linear coding has been standardized as application-layer forward error correction (AL-FEC) for multimedia broadcast and multicast services (MBMS)~\cite{threeGPP2005}. Thus, the efficiency of a software-based implementation will depend on the underlying hardware architecture, e.g., ARM or x86, and on the chosen programming language, that is, its capability to directly access hardware resources and its flexibility to explore parallelization.



\bibliographystyle{IEEEtran}
\bibliography{IEEEabrv, references}

\end{document}